\documentclass[fleqn,usenatbib]{mnras}

\usepackage{newtxtext,newtxmath}

\usepackage[T1]{fontenc}

\DeclareRobustCommand{\VAN}[3]{#2}
\let\VANthebibliography\thebibliography
\def\thebibliography{\DeclareRobustCommand{\VAN}[3]{##3}\VANthebibliography}

\usepackage{graphicx}	
\usepackage{amsmath}	
\usepackage[normalem]{ulem}	
\usepackage{textcomp}	
\usepackage{orcidlink}
\usepackage[nolist]{acronym}
\usepackage{cuted}
\hypersetup{urlcolor=violet}

\defcitealias{Casarini2016}{C16}

\newacro{BAO}{baryon acoustic oscillation}
\newacro{BDM}{bound density maxima}
\newacro{CDM}{cold dark matter}
\newacro{CIC}{cloud-in-cell}
\newacro{CMB}{cosmic microwave background}
\newacro{CPL}{Chevallier--Polarski--Linder}
\newacro{DE}{dark energy}
\newacro{DESI}{Dark Energy Spectroscopic Instrument}
\newacro{EoS}{equation of state}
\newacro{GR}{general relativity}
\newacro{PDF}{probability distribution function}
\newacro{PM}{particle mesh}

\title[Reproducing $w_0$-$w_a$ at the field level with fixed $w$]{Using fixed $w$ to reproduce evolving $w_0$-$w_a$ dark energy at the field level}

\author[Y.~Meng et al.]{
Yuxi Meng\orcidlink{0009-0004-6895-6743}$^{1, 2}$\thanks{E-mail: yxmeng@bao.ac.cn},
Daniela Saadeh\orcidlink{0000-0001-7086-4216}$^{2}$,
Cheng-Zong Ruan\orcidlink{0000-0001-7255-0536}$^{2}$,
Baojiu Li\orcidlink{0000-0002-1098-9188}$^{2,1}$,
and Kazuya Koyama\orcidlink{0000-0001-6727-6915}$^{3,4,5}$
\\
$^{1}$National Astronomical Observatories, Chinese Academy of Sciences, Datun Road, Beijing 100101, China\\
$^{2}$Institute for Computational Cosmology, Department of Physics, Durham University, South Road, Durham DH1 3LE, UK\\
$^{3}$Institute of Cosmology \& Gravitation, University of Portsmouth, Dennis Sciama Building, Burnaby Road, Portsmouth PO1 3FX, UK\\
$^{4}$Kavli IPMU (WPI), UTIAS, The University of Tokyo, Kashiwa, Chiba 277-8583, Japan\\
$^{5}$Yukawa Institute for Theoretical Physics, Kyoto University, Kitashirakawa Oiwakecho, Sakyo-ku, Kyoto 606-8502, Japan
}

\date{Accepted XXX. Received YYY; in original form ZZZ}

\pubyear{\the\year{}}

\begin{document}
\label{firstpage}
\pagerange{\pageref{firstpage}--\pageref{lastpage}}
\maketitle


\begin{abstract}
Recent results from the Dark Energy Spectroscopic Instrument (DESI) hint at dynamical dark energy, making the accurate modelling of such cosmologies increasingly important for future analyses. In this work, we show that because, when written in superconformal time $\tau$, the $N$-body equations of motion depend on the background cosmology only through the expansion history $a(\tau)$, matching $a(\tau)$ allows us to map one cosmology onto another. As a first, observationally relevant, application of this new method, we match a Chevallier--Polarski--Linder (CPL) cosmology to a constant-$w$ cosmology with the closest $a(\tau)$. This allows us to use a single fixed-$w$ simulation to reproduce the simulation predictions of the target CPL cosmology at the field level at redshifts $0\le z\le3$. We validate this matching with 100 pairs of CPL and fixed-$w$ \textsc{mg-glam} simulations, spanning $0.1\le\Omega_{\rm m}\le0.5$, $-1.4\le w_0\le-0.5$ and $-1.8\le w_a\le-0.15$. For a DESI-like cosmology, the matter power spectra of the matched and target simulations agree to 0.24\% for $k\lesssim1\,h\,{\rm Mpc}^{-1}$ and 0.8\% up to $k\lesssim3\,h\,{\rm Mpc}^{-1}$ at $0\le z\le3$, and the density phases are reproduced with a stochasticity below 0.02\%. Displacements, velocities, momenta, redshift-space multipoles, bispectra and the halo and void abundances are similarly well predicted. For most models we consider, we achieve subpercent error in the power spectrum up to scales $k\lesssim1\,h\,{\rm Mpc}^{-1}$ at all redshifts, with density stochasticity below 0.03\% over the same scales. The method allows constant-$w$ simulations and field-level emulators to be used for CPL cosmologies, and generalises to any cosmology with no additional scale dependence.

\end{abstract}




\section{Introduction}

The nature of \ac{DE} has long been a central question in cosmology. Recent \ac{BAO} measurements from the \ac{DESI}, combined with the \ac{CMB} and type Ia supernovae, point to a time-evolving dark energy over a cosmological constant \citep{DESI2025}. This preference is expressed in the \ac{CPL} parametrisation, in which the dark energy \ac{EoS} parameter $w$ evolves with the scale factor $a$ as $w(a)=w_0+w_a(1-a)$. This form captures the expansion histories of a wide range of physical dark energy models \citep{Chevallier2001,Linder2003}. Cosmological analyses of current and upcoming surveys therefore need to explore an extended parameter space that includes $w_0$ and $w_a$.

Extending the parameter space comes at a cost on nonlinear scales, where predictions rely on $N$-body simulations. These are the standard tool to follow the nonlinear growth of structure and underpin most analyses on small scales, but are too expensive to run for every model that must be sampled in a cosmological data analysis. Emulators trained on suites of $N$-body simulations have emerged as an efficient alternative for predicting nonlinear statistics \citep[e.g.][]{Heitmann2009,Lawrence2017,DeRose2019,Nishimichi2019,Angulo2021,Knabenhans2021}, and are widely used in cosmological inference. For dynamical dark energy, some emulators predict the nonlinear matter power spectrum directly as a function of $w_0$ and $w_a$ \citep[][]{Knabenhans2021,Yang2025}.

Another approach is to reuse existing simulations or emulators by mapping between different cosmologies. Rescaling methods can take a simulation run in one cosmology and relabel its length, mass and time units to approximate other cosmologies \citep{Angulo2010}, including modified gravity models \citep{Mead2015}. Evolution mapping instead relates cosmologies at equal linear clustering amplitude \citep{Sanchez2022,Esposito2024}. For dynamical dark energy, \citet[hereafter \citetalias{Casarini2016}]{Casarini2016} matched the target model, at each redshift, to a constant-$w$ model sharing its comoving distance to the last-scattering surface and its $\sigma_8(z)$, a method previously introduced by \citet{Casarini2009} \citep[see also][]{Linder2005,Francis2007}. Using this matching, \citetalias{Casarini2016} extended the Coyote emulator \citep{Heitmann2010,Heitmann2009,Lawrence2010,Heitmann2014}
to $w_0w_a$ models, showing, by means of $N$-body simulations, that it reproduces the nonlinear matter power spectrum to better than 0.5\% at $k<2\,h\,{\rm Mpc}^{-1}$ and 1\% at $k<3\,h\,{\rm Mpc}^{-1}$ for $0\le z\le3$. \citet{Chen2026} applied the same matching to models favoured by \ac{DESI}, achieving an accuracy of about 1\%. These methods have been validated on summary statistics like the matter power spectrum or the halo mass function, but agreement at the field level is unknown.

Summary statistics, however, can not capture all the information in the nonlinear field. Field-level emulators go one step further and predict the outputs of $N$-body simulations at the field level, either the density field \citep[e.g.][]{Kang2026} or the full particle output. The latter use neural networks to map the linear displacement field to the nonlinear displacements and velocities of $N$-body simulations with percent-level accuracy \citep{Jamieson2023,Scoggins2025}, with redshift dependence through the linear growth factor \citep{Jamieson2025}. They have also been applied to beyond $\Lambda$CDM models, such as modified gravity \citep{Saadeh2025a,Saadeh2025b}. Other approaches combine an emulator with particle-mesh force evaluations \citep{Bartlett2025} or use neural cellular automata \citep{Jacobus2026}. These approaches are deterministic, mapping each input field to a single output. Generative models, such as conditional generative adversarial networks, have also been applied to map $\Lambda$CDM fields to modified gravity \citep{Gondhalekar2025}.

A key advantage of field-level emulators is that they can provide full $N$-body snapshots, so that any summary statistic, including the ones in redshift space, can then be measured from their output. They enable inference and analysis at the field level, such as in the reconstruction of the initial conditions \citep{Doeser2024} and the painting of haloes, galaxies and other tracers on fast dark matter fields \citep{Pandey2025,Scelfo2026}, including in modified gravity \citep{GarciaFarieta2024}. However, none of them currently covers $w_0w_a$ cosmologies. \citet{Kaushal2022} have tested a field-level emulator with $w\neq-1$, but only for constant $w$. Extending these emulators to CPL models would require training on new simulation suites.

\begin{figure*}
    \centering
    \includegraphics[width=1\linewidth]{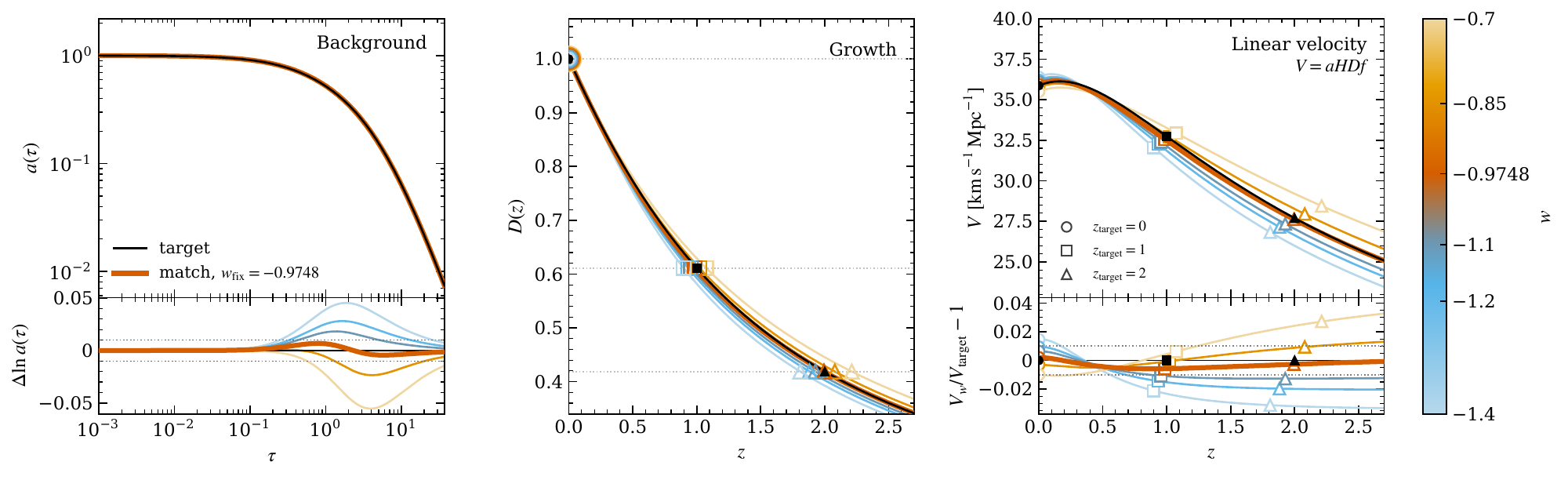}
    \caption{Matching procedure for the DESI-like target $(\Omega_{\rm m},w_0,w_a)=(0.3191,-0.752,-0.86)$. Left: $a(\tau)$ of the target (black) and of constant-$w$ models (thin coloured lines, see colour bar), with $\Delta\ln a(\tau)$ (Eq.~\eqref{eq:J}) in the lower panel. The thick line is the match, $w_{\rm fix}=-0.9748$, with $\epsilon=0.0029$. Middle: linear growth factor $D(z)$, normalised to $D(0)=1$. For $z_{\rm target}=0$, 1 and 2, open markers indicate the redshift at which each constant-$w$ model reaches the target value of $D$ (filled black markers). Right: linear velocity prefactor $V=aHDf$ before rescaling, with $V_w/V_{\rm target}-1$ in the lower panel. Each curve is plotted against its own redshift, and the markers indicate the matched epochs. For the match, $V$ differs from the target by at most 0.6\% at the matched epochs.}
    \label{fig:matching}
\end{figure*}

In this work, we propose a new matching method that is based on a universality of the $N$-body equations for models that do not introduce additional length scales, including $\Lambda$CDM and most non-clustering dark energy models. The key idea is to match the expansion history in superconformal time, defined by ${\rm d}\tau\propto{\rm d}t/a^2$. With this time variable, it can be shown that the $N$-body equations of motion depend on the background only through $\Omega_{\rm m}a(\tau)$ \citep{Martel1998}, so that two models with the same $\Omega_{\rm m}a(\tau)$ and the same initial conditions produce exactly identical particle trajectories. 

Here, we apply this general method to evolving dark-energy models with a CPL equation of state, showing it is possible to reproduce them at the field level by means of a constant \ac{EoS} chosen to match $a(\tau)$, while keeping $\Omega_{\rm m}$ the same. By matching the full expansion history, we are able to reproduce snapshots at \textit{any} time from the initial redshift to $z=0$ with a \textit{single} matching, which is the key difference from the earlier work of \citetalias{Casarini2016}. To achieve this, $a(\tau)$ is matched over the expansion history up to the present day, with all other cosmological parameters kept fixed\footnote{We note that neither of these is necessary. In a separate work we will show that this method also works for other models, such as scale-independent modified gravity, where $\tau$ in the matched model is shifted or rescaled compared with $\tau$ in the target model, so that the matched simulation may need to stop at $z>0$ or $z<0$. Furthermore, for more complicated models than vanilla \ac{CPL}, the matched model may need to have different cosmological parameters.}.

We validate the new mapping method with 100 pairs of target and matched simulations with $1024^3$ particles and the same initial phases, on a grid of $(\Omega_{\rm m},w_0,w_a)$ that includes the region favoured by DESI. Beyond the power spectrum, we test the density field at the field level through its stochasticity, and compare the Lagrangian displacement, Lagrangian velocity, Eulerian momentum, redshift-space distortions, bispectrum, haloes and voids. We also compare our matching with that of \citetalias{Casarini2016} and test the dependence on the $N$-body code. We show that this matching method gives sub-percent accuracy in its predictions of all tested summary statistics for current observationally-relevant CPL cosmologies. A constant-$w$ simulation suite or field-level emulator can therefore be used for CPL cosmologies without retraining.

This paper is structured as follows. Section~\ref{sec:methods} describes the method underlying this work, i.e.\ the matching in superconformal time, and the $N$-body simulations we use to validate it. Section~\ref{sec:results} tests our matching against a range of summary statistics, including the density power spectrum, bispectrum and stochasticity, the power spectra of the Lagrangian displacement, Lagrangian velocity and Eulerian momentum, redshift-space multipoles, and halo and void abundances. We then compare our method against the original work of \citetalias{Casarini2016}, and examine the limitations of matching to a single constant $w$ by stress-testing the method with a positive $w_a$, i.e.\ $w_a=+0.4$, with 25 further simulation pairs. For these targets, dark energy is still important at early times, which a constant $w$ cannot reproduce. Section~\ref{sec:conclusions} summarises our conclusions. In the Appendices, we first show that our results hold for two other $N$-body codes, \textsc{pkdgrav3} and \textsc{mp-gadget} (Appendix~\ref{app:codes}), and then show the effect of rescaling the velocities of the match using linear theory (Appendix~\ref{app:velcor}).

Throughout this paper we adopt the common conventions in the literature: $\Omega_{\rm m}, \Omega_{\rm r}, \Omega_{\rm b}$ denote the present-day density parameters for matter, radiation and baryons respectively, $H_0$ is the Hubble constant with $h=H_0/\left(100\,{\rm km}\,{\rm s}^{-1}\,{\rm Mpc}^{-1}\right)$, $\sigma_8$ is the root-mean-square density fluctuation at $z=0$ smoothed at $8~h^{-1}{\rm Mpc}$, $n_{\rm s}$, $A_{\rm s}$ are the power index and amplitude of the primordial power spectrum, and $a$ is the scale factor normalised to $a=1$ at redshift $z=0$ (today).

\section{Methods}\label{sec:methods}
\subsection{Superconformal Time and Matching}\label{sec:matching}

We consider spatially flat cosmologies containing matter and dark energy. In the following, we model dark energy as a barotropic fluid with \ac{EoS} $P_{\rm DE}=w\rho_{\rm DE}c^2$, where $P_{\rm DE}$ and $\rho_{\rm DE}$ are the \ac{DE} pressure and energy density, respectively, and $c$ the speed of light. For $w=-1$, we recover a cosmological constant. 

For each target CPL cosmology, we construct a matched cosmology with a constant \ac{EoS} parameter $w_{\rm fix}$ that shares all other cosmological parameters with the target. The matching proceeds in three steps: 
\begin{enumerate}
\item[(i)] choose the $w_{\rm fix}$ whose expansion history is closest to that of the target,
\item[(ii)] find, for each target redshift, the matched redshift with an equal linear growth factor $D$, and
\item[(iii)] rescale the velocities and redshift-space displacements based on linear theory.
\end{enumerate}
We describe each step below. 

The equations of an $N$-body system take a particularly simple form when expressed in superconformal time \citep{Martel1998}. We use the dimensionless look-back superconformal time
\begin{equation}\label{eq:tau}
\tau(a)\equiv H_0\int_{t(a)}^{t_0}\frac{\mathrm{d}t'}{a^2(t')}
=\int_a^1\frac{\mathrm{d}\ln a'}{a'^2 E(a')},
\end{equation}
where $t$ is the cosmic time, $t_0$ its present-day value, $H(a)$ the Hubble parameter, whose present-day value is $H_0$ (the Hubble constant), and $E(a)\equiv H(a)/H_0$ the normalised expansion rate. For a flat universe with CPL dark energy \citep{Chevallier2001,Linder2003},
\begin{equation}
    w(a)=w_0+w_a(1-a),
\end{equation}
we have
\begin{equation}\label{eq:Ea}
E^2(a)=\Omega_{\rm m}a^{-3}+(1-\Omega_{\rm m})\,a^{-3(1+w_0+w_a)}\,e^{-3w_a(1-a)},
\end{equation}
where $\Omega_{\rm m}$ is the present-day matter density parameter. For simplicity and without loss of generality, we neglect radiation in $E(a)$, as in our $N$-body simulations, so $1-\Omega_{\rm m}$ is the dark energy density. A constant-$w$ model corresponds to $w_0=w$, $w_a=0$.  Using $\mathrm{d}\tau=-H_0\,\mathrm{d}t/a^2$, the equations of motion become
\begin{equation}\label{eq:eom}
\left\{
\begin{aligned}
\frac{\mathrm{d}^2\boldsymbol{x}}{\mathrm{d}\tau^2}&=-\boldsymbol{\nabla}_x{\Phi},\\
\nabla_x^2{\Phi}&=\frac{3}{2}\,\Omega_{\rm m}\,a(\tau)\,\delta,
\end{aligned}
\right.
\end{equation}
where $\boldsymbol{x}$ is the comoving position, $\Phi\equiv a^2\phi/H_0^2$ with $\phi$ the peculiar gravitational potential, and $\delta$ is the matter overdensity. Note that, in Eq.~\eqref{eq:eom}, the cosmology enters only through $\Omega_{\rm m}a(\tau)$. With $\Omega_{\rm m}$ fixed, two models that start from the same initial conditions and have identical $a(\tau)$ would therefore produce identical particle trajectories. The Hubble constant $H_0$ does not appear in the equations explicitly, but is implicitly included in the units of time ($\tau$) and space. We adopt the same $H_0$, and hence $h=H_0/(100\,{\rm km\,s^{-1}\,Mpc^{-1}})$, for the target and the matched models, so that both are expressed in the same units of time, length and momentum and their outputs can be compared directly.

Although the target and matched models share $\Omega_{\rm m}$ and $h$, a constant $w$ cannot reproduce the $a(\tau)$ of a CPL model exactly. For each target, we therefore choose the constant-$w$ model whose $a(\tau)$ is closest to that of the target CPL (Fig.~\ref{fig:matching}, left panel). In practice, we compare the two expansion histories, $a(\tau)$, on a grid of $N=256$ logarithmically-spaced values of $\tau$ in the range $10^{-3}\le\tau\le37$, which emphasises the late-time evolution. We therefore define the matching error as
\begin{equation}\label{eq:J}
{\epsilon}(w)\equiv\sqrt{\frac{1}{N}\sum_{i=1}^{N}\left[\Delta\ln a(\tau_i)\right]^2},
\end{equation}
where $\Delta\ln a(\tau)\equiv\ln a_w(\tau)-\ln a_{\rm target}(\tau)$, $a_w(\tau)$ is the expansion history of a constant-$w$ model, and $w_{\rm fix}$ is the value of $w$ that minimises $\epsilon$. Unlike the relative difference $a_w/a_{\rm target}-1$, the log difference here is symmetric between a matched scale factor that is too large and one that is too small by the same factor. Our conclusions are not sensitive to the particular number of points $N$ or to the $\tau$ range.

The method described above with Eq.~\eqref{eq:eom} is \textit{exact}, in the sense that if the target and matched cosmologies have identical $a(\tau)$, then their simulation predictions are completely identical, and there is no approximation involved apart from those used to derive the equations from \ac{GR}, and the assumption that \ac{DE} clustering is negligible. In practice, however, as we have noticed from Fig.~\ref{fig:matching}, the two models do not have identical $a(\tau)$, so that there is a slight mismatch between the structure growths for the same $\tau$ or $a$.

Rather than using a different constant-$w$ model at each redshift, as in \citetalias{Casarini2016}, we use a single $w_{\rm fix}$ per target, trading some accuracy for efficiency. As a result, each target redshift $z_{\rm target}$ requires a matched redshift $z_{\rm match}$ in the matched cosmology at which the growth difference is minimised. The two models share all other parameters, including $n_{\rm s}$ and the present-day $\sigma_8$. Their initial conditions have the same phases but slightly different amplitudes\footnote{This is because we generate the initial conditions separately for the target and matched cosmologies, using their respective linear power spectra at $z_{\rm init}=100$, but require that they have the same $\sigma_8$ today. This is a choice, on which we will comment further in the discussion and conclusions.}. For each $z_{\rm target}$, we can therefore find the $z_{\rm match}$ that satisfies $D_{\rm match}(z_{\rm match})=D_{\rm target}(z_{\rm target})$, with $D$ normalised to unity at $z=0$ in each model (Fig.~\ref{fig:matching}, middle panel). At this redshift, the two models have the same linear amplitude, and therefore the same linear density field. This is where the efficiency comes from: since any target redshift can be matched in this way, a single matched simulation serves all target redshifts, whereas matching a different constant-$w$ model at each redshift would require a separate $N$-body simulation for every target redshift. This is similar in spirit to evolution mapping \citep{Sanchez2022}. In the following, we label each pair of matched outputs by its target redshift and write $z\equiv z_{\rm target}$.

Matching $D$ is not sufficient to guarantee that the velocities will match too. In fact, in linear theory, the peculiar velocity is
\begin{equation}\label{eq:vel}
\boldsymbol{v}(\boldsymbol{q},z)=a(z)H(z)f(z)D(z)\,\boldsymbol{\Psi}_0(\boldsymbol{q}),
\end{equation}
where $f\equiv\mathrm{d}\ln D/\mathrm{d}\ln a$ is the linear growth rate, $\boldsymbol{q}$ the initial position of a particle and $\boldsymbol{\Psi}_0$ the linear displacement field with $D=1$. At matched epochs, $D$ is equal, but the factor $aHf$ may not be (Fig.~\ref{fig:matching}, right panel). Following \citet{Esposito2024}, we therefore rescale the velocities of the matched simulation by
\begin{equation}\label{eq:velcorr}
\boldsymbol{v}_{\rm match}\rightarrow\frac{(aHf)_{\rm target}}{(aHf)_{\rm match}}\,\boldsymbol{v}_{\rm match},
\end{equation}
where the target and matched quantities are evaluated at $z_{\rm target}$ and $z_{\rm match}$, respectively. 
For the velocity and momentum fields (Sect.~\ref{sec:fields}), the correction is a constant
rescaling at each output, independent of position and scale, which changes only the field amplitude, i.e.\ the power spectrum, and leaves the stochasticity unchanged. The redshift-space positions are then obtained from the rescaled velocities as $\boldsymbol{s}=\boldsymbol{x}_{\rm match}+v_{{\rm match},\parallel}/(a_{\rm target}H_{\rm target})\,\hat{\boldsymbol{n}}$, where $\hat{\boldsymbol{n}}$ is the line-of-sight direction and $v_{{\rm match},\parallel}$ is the line-of-sight component of the rescaled velocity. The conversion uses the $aH$ of the target simulation at $z_{\rm target}$, so the redshift-space positions are those of the target model. Unlike for the velocity and momentum fields, the rescaling therefore changes the redshift-space density field itself, not only its amplitude. We show the effect of rescaling in Appendix~\ref{app:velcor}.

\subsection{Simulations} \label{sec:simulations}

We run our simulations with \textsc{mg-glam}, a \ac{PM} $N$-body code \citep{Klypin2018,Ruan2022,HernandezAguayo2022}, in its GR mode, i.e.\ without any modified-gravity force. We modify the code to (i) include the CPL equation of state, (ii) write snapshots at exactly the requested redshifts instead of the nearest time step, and (iii) output velocities at the same time as the positions, rather than half a time step apart.

Each simulation follows $N_{\rm p}=1024^3$ particles in a box of side length $L=1000\,h^{-1}{\rm Mpc}$, giving a mass resolution of $m_{\rm p}=7.75\times10^{10}\,(\Omega_{\rm m}/0.3)\,h^{-1}M_\odot$. Forces are computed on a $4^3\times N_{\rm p}=4096^3$ mesh, so the force resolution is set by the cell size, $244.14\,h^{-1}{\rm kpc}$. All simulations start at $z_{\rm init}=100$ with an initial step $\Delta a_{\rm init}=4\times10^{-4}$, which the code increases as the simulation advances \citep[see][for details]{Klypin2018}, giving 158--162 time steps in total. 

The linear power spectra are computed using \textsc{camb}\footnote{\url{https://camb.info}}
(\citealt{Lewis2000,Howlett2012,Lewis2026}), using the parameterised post-Friedmann prescription to handle phantom crossing \citep{Fang2008}. We include radiation by setting $\Omega_{\rm r}=9.1\times10^{-5}$. The initial conditions for all simulations are generated using second-order Lagrangian perturbation theory (2LPT, \citealt{Scoccimarro1998}), in the implementation of the \textsc{fml} library \citep{Fiorini2021}\footnote{\url{https://github.com/HAWinther/FML}}, which we extended to support a user-defined growth history, set here to that of the \ac{CPL} background.

The initial conditions displace the particles from a uniform grid of $1024^3$ sites. We label each particle by its grid site $\boldsymbol{q}$, its Lagrangian coordinate, so that its comoving position is $\boldsymbol{x}(\boldsymbol{q})=\boldsymbol{q}+\boldsymbol{\Psi}(\boldsymbol{q})$, where $\boldsymbol{\Psi}$ is the Lagrangian displacement field. The Lagrangian velocity field, $\boldsymbol{v}(\boldsymbol{q})$, is the velocity of the particle labelled by $\boldsymbol{q}$. Eulerian fields are instead defined at comoving positions $\boldsymbol{x}$, for example the density and momentum fields. We obtain the overdensity, $\delta(\boldsymbol{x})\equiv[\rho(\boldsymbol{x})-\bar\rho]/\bar\rho$, where $\bar\rho$ is the mean density in the box, and the momentum, $\boldsymbol{p}(\boldsymbol{x})=[1+\delta(\boldsymbol{x})]\,\boldsymbol{v}(\boldsymbol{x})$, by assigning the particles and their velocities to the $1024^3$ mesh with the \ac{CIC} scheme.

Motivated by the recent \ac{DESI} results \citep{DESI2025}, we construct a grid of target cosmologies with $\Omega_{\rm m}=(0.1,0.2,0.3,0.4,0.5)$, $w_0=(-1.4,-1.175,-0.95,-0.725,-0.5)$ and $w_a=(-1.8,-1.25,-0.7,-0.15)$, i.e.\ 100 targets, together with 25 targets with $w_a=+0.4$ as a stress test for later, giving 125 in total. All models are spatially flat. The remaining parameters follow Planck 2015 \citep{Planck2016}: $\Omega_{\rm b}=0.0486$, $n_{\rm s}=0.9667$, $\sigma_8=0.8159$ and $h=0.6774$. Each target uses a different random seed, while the target and its match share the same seed, giving 125 independent realisations. As a representative example, we also adopt a \ac{DESI}-like cosmology, $(\Omega_{\rm m},w_0,w_a)=(0.3191,-0.752,-0.86)$, the posterior mean for flat $w_0w_a$CDM from \ac{DESI}+\ac{CMB}+DESY5 \citep[Table V]{DESI2025}, with all other parameters as in the grid. It is closer to current constraints than the grid points, and allows a direct comparison with other studies that adopt the \ac{DESI} results.

We identify haloes with the \ac{BDM} halo finder \citep{Klypin1997,Knebe2011}, which locates density maxima, removes unbound particles and defines haloes as spherical overdensities. We use the bound mass $M_{200{\rm m}}$, enclosed within a radius where the mean density is $200$ times the mean matter density. Voids are identified with the spherical void finder of \textsc{pylians3}\footnote{\url{https://github.com/franciscovillaescusa/Pylians3}} \citep{VillaescusaNavarro2018}, which implements the algorithm of \citet{Banerjee2016}: it applies spherical top-hat filters of decreasing radius, from 125 to $3.9\,h^{-1}{\rm Mpc}$, to the \ac{CIC}
density field and keeps non-overlapping spheres whose mean overdensity is below a threshold $\delta_{\rm th}$, which we set to $-0.7$. 
\section{Results}\label{sec:results}

We now compare the target and matched simulations. We start with the density field in Sect.~\ref{sec:density}, first for the \ac{DESI}-like pair and then for the grid of models spanning $(\Omega_{\rm m},w_0,w_a)$. We assess our matching by comparing density slices and summary statistics of the density field, including the one-point distribution function, the power spectrum, the stochasticity and the bispectrum. Section~\ref{sec:fields} extends the comparison beyond the real-space density field, to the Lagrangian displacement and velocity, the Eulerian momentum and the redshift-space multipoles. In Sect.~\ref{sec:haloes}, we examine halo and void abundances. We then compare our matching with the redshift-dependent matching of \citetalias{Casarini2016} in Sect.~\ref{sec:se}. In Sect.~\ref{sec:extreme}, we use the simulation pairs with $w_a=+0.4$ to test the limits of matching a \ac{CPL} model with a single constant $w$.

In Appendix~\ref{app:codes}, we repeat the comparison for the \ac{DESI}-like pair with two other $N$-body codes, \textsc{pkdgrav3} and \textsc{mp-gadget}, demonstrating that the power spectrum errors of the matching do not depend on the code.

Unless stated otherwise, the results for the grid use the 100 simulation pairs with $w_a<0$, and the velocities of the matched simulations are rescaled with Eq.~\eqref{eq:velcorr}. Appendix~\ref{app:velcor} gives further details of the effect of this rescaling.

\begin{figure*}
    \centering
    \includegraphics[width=1\linewidth]{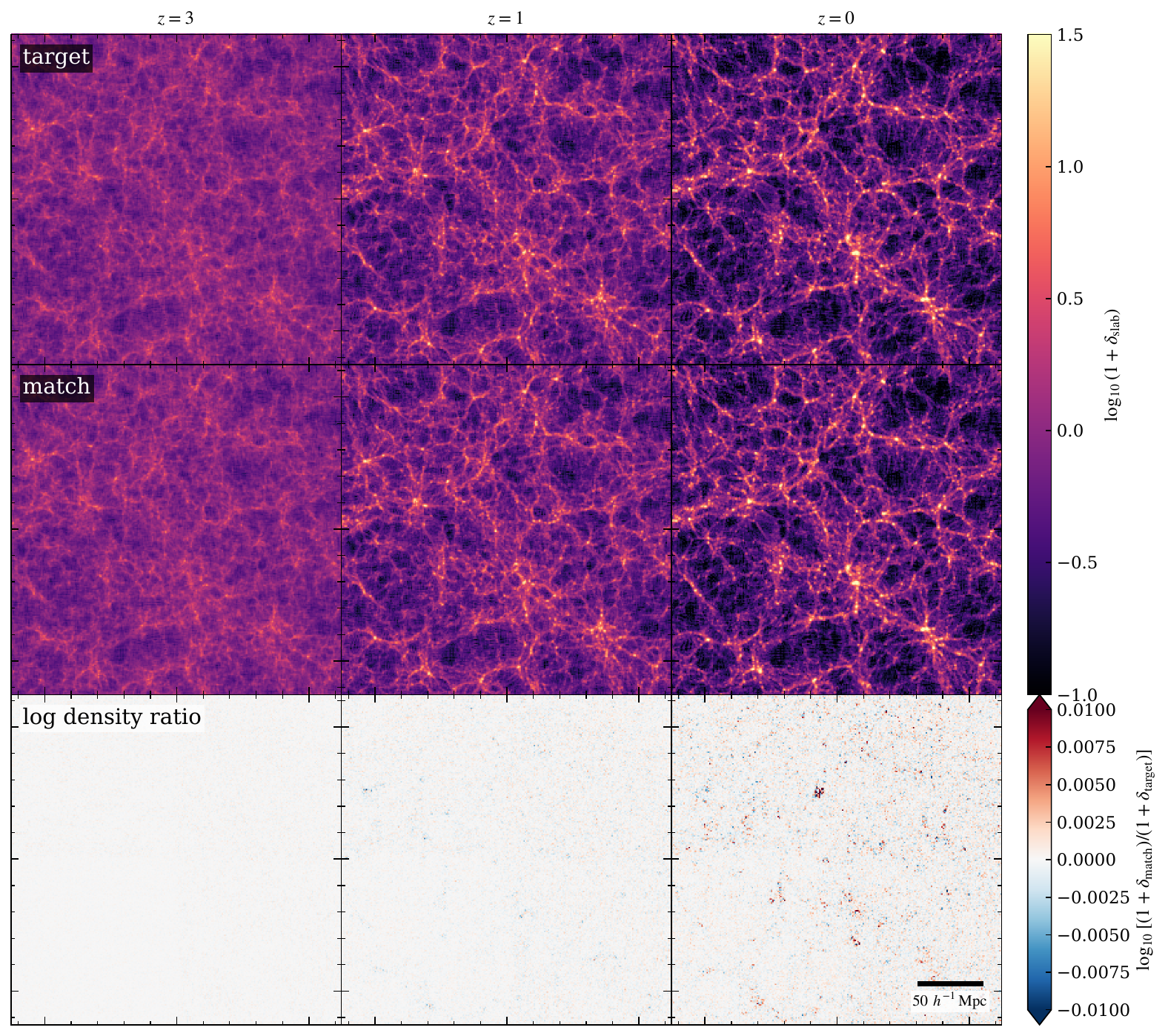}
    \caption{Density slices for the DESI-like target $(\Omega_{\rm m},w_0,w_a)=(0.3191,-0.752,-0.86)$ and its match ($w_{\rm fix}=-0.9748$) at $z=3$, 1 and 0 (columns). Rows: target, match, and the log density ratio $\log_{10}[(1+\delta_{\rm match})/(1+\delta_{\rm target})]$. Each panel covers the same $250\times250\,(h^{-1}{\rm Mpc})^2$ region with a projection depth of $20.5\,h^{-1}{\rm Mpc}$. The differences grow with time, but at all three redshifts at least 99.1\% of the pixels agree with the target to within 1\%.}
    \label{fig:slices}
\end{figure*}

\begin{figure}
    \centering
    \includegraphics[width=1\linewidth]{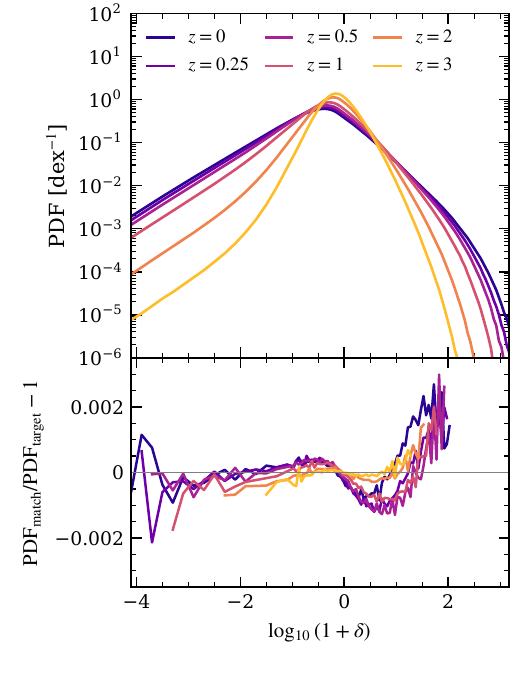}
    \caption{The \acf{PDF} of $\log_{10}(1+\delta)$ for the \ac{DESI}-like target and matched pair, measured on the full $1024^3$ CIC mesh. Top: \ac{PDF} of the target. The \ac{PDF} of the match is indistinguishable and is not shown. Bottom: relative difference between the matched and target \ac{PDF}s, with deviations below 0.3\% at all six outputs. Colours denote the redshift $z$, as indicated by the legend.}
    \label{fig:pdf}
\end{figure}

\begin{figure}
    \centering
    \includegraphics[width=1\linewidth]{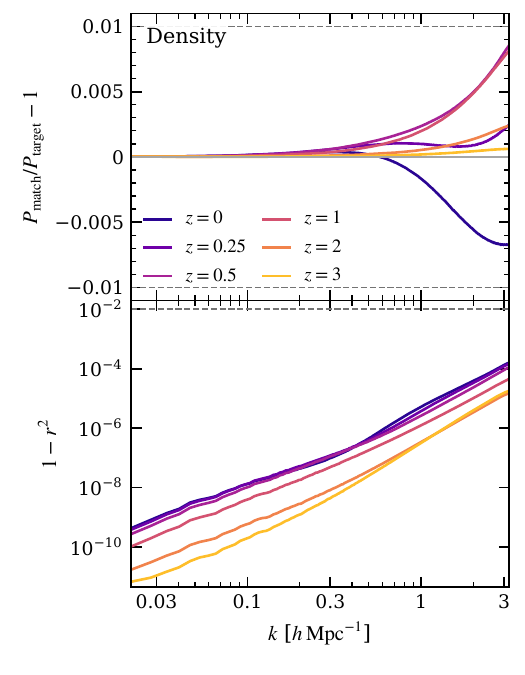}
    \caption{Power spectrum error $\Delta P/P$ (top) and stochasticity $1-r^2$ (bottom) of the density field for the \ac{DESI}-like (DESI posterior mean) target and matched pair at six target redshifts. Dashed lines mark $\pm1\%$ (top) and $10^{-2}$ (bottom). The spectrum and field-level statistics show good agreement between target and match.}
    \label{fig:power}
\end{figure}

\subsection{The density field statistics}
\label{sec:density}

We assign the particles to a $1024^3$ mesh using the \ac{CIC} scheme, to obtain the density field $\rho(\boldsymbol{x})$ and the overdensity $\delta(\boldsymbol{x})\equiv\rho(\boldsymbol{x})/\bar\rho-1$, where $\bar\rho$ is the mean density in the box. We compare the target and matched density fields at the spectrum and field levels.

We first examine the \ac{DESI}-like target cosmology $(\Omega_{\rm m},w_0,w_a)=(0.3191,-0.752,-0.86)$, matched by $w_{\rm fix}=-0.9748$. For this pair, the six outputs at $z=0$, 0.25, 0.5, 1, 2 and 3 correspond to $z_{\rm match}=0$, 0.245, 0.490, 0.985, 1.99 and 3.01. In Figure~\ref{fig:slices}, we show sample density slices, $250\times250\,(h^{-1}{\rm Mpc})^2$ wide and $20.5\,h^{-1}{\rm Mpc}$ thick, through the centre of the box, of the target and matched simulations, which are visually indistinguishable. 
The differences grow with time, but even at $z=0$ the projected densities of target and match agree to within 1\% in 99.1\% of the pixels, and the few larger differences lie preferentially near dense structures.

The one-point \ac{PDF}s of $\log_{10}(1+\delta)$ agree to $<0.3\%$ at all six outputs (Fig.~\ref{fig:pdf}), with the largest deviations at the low- and high-density ends of the distribution, where the bins contain fewer mesh cells and are therefore noisier.

The matter power spectrum $P(k)$ is defined by $\langle\delta(\boldsymbol{k})\,\delta^*(\boldsymbol{k}')\rangle=(2\pi)^3\delta_{\rm D}(\boldsymbol{k}-\boldsymbol{k}')\,P(k)$, where $\delta(\boldsymbol{k})$ is the Fourier transform of the density contrast $\delta(\boldsymbol{x})$, and $\delta_{\rm D}$ is the Dirac $\delta$ function. We measure $P(k)$ on the same $1024^3$ \ac{CIC} mesh, correcting for the \ac{CIC} window function, in bins of width $k_{\rm F}=2\pi/L$, and do not subtract shot noise. We quantify the difference using $\Delta P/P\equiv P_{\rm match}(k)/P_{\rm target}(k)-1$, where $P_{\rm match}$ is measured at $z_{\rm match}$ and $P_{\rm target}$ at $z_{\rm target}$.

\begin{figure*}
    \centering
    \includegraphics[width=1\linewidth]{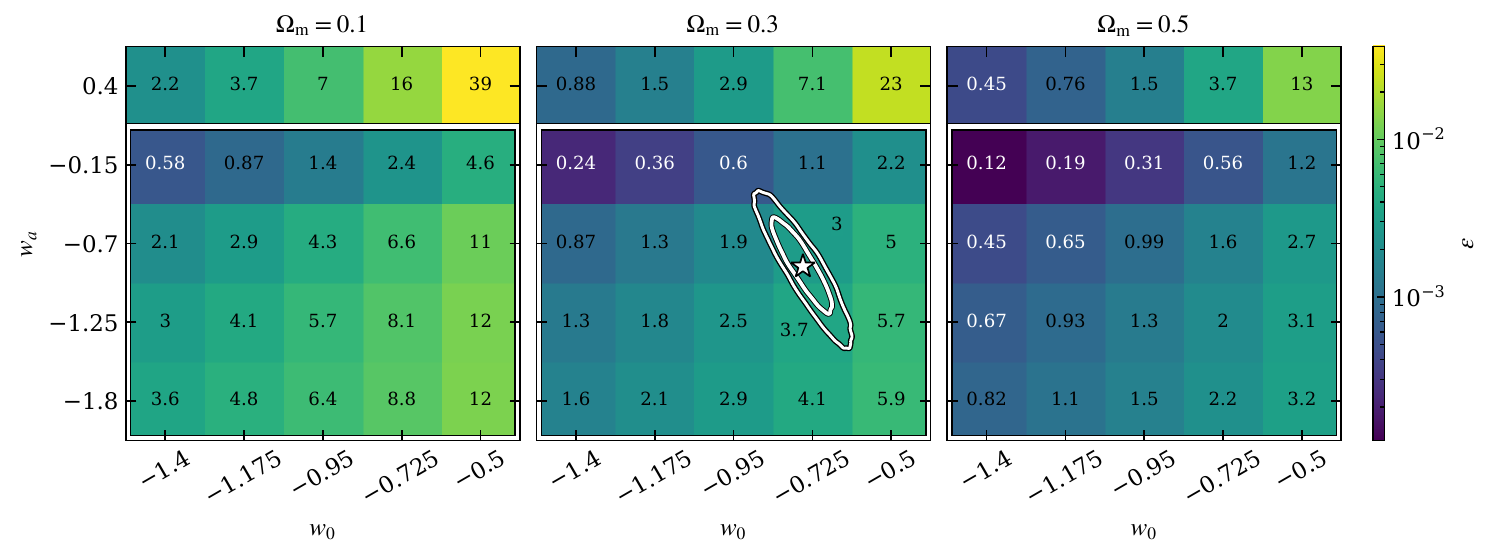}
    \caption{Matching error $\epsilon$ over the \ac{CPL} parameter grid. The three panels fix $\Omega_{\rm m}=0.1$, 0.3 and 0.5. The axes are $w_0$ and $w_a$. Numbers give $10^3\epsilon$ and colours $\epsilon$, with the same colour scale as Fig.~\ref{fig:ensemble}. The white frame marks the $w_a<0$ models used in the main analysis. On the middle panel, the white contours show the 68\% and 95\% confidence regions of the \ac{DESI} DR2 constraint from \ac{DESI}+\ac{CMB}+DESY5, marginalised over $\Omega_{\rm m}$, and the star marks its posterior mean. $\epsilon$ generally increases with $w_0$ and $|w_a|$, and is largest in the upper-right corner of the left panel, $(\Omega_{\rm m},w_0,w_a)=(0.1,-0.5,0.4)$.}
    \label{fig:Jmap}
\end{figure*}

\begin{figure*}
    \centering
    \includegraphics[width=1\linewidth]{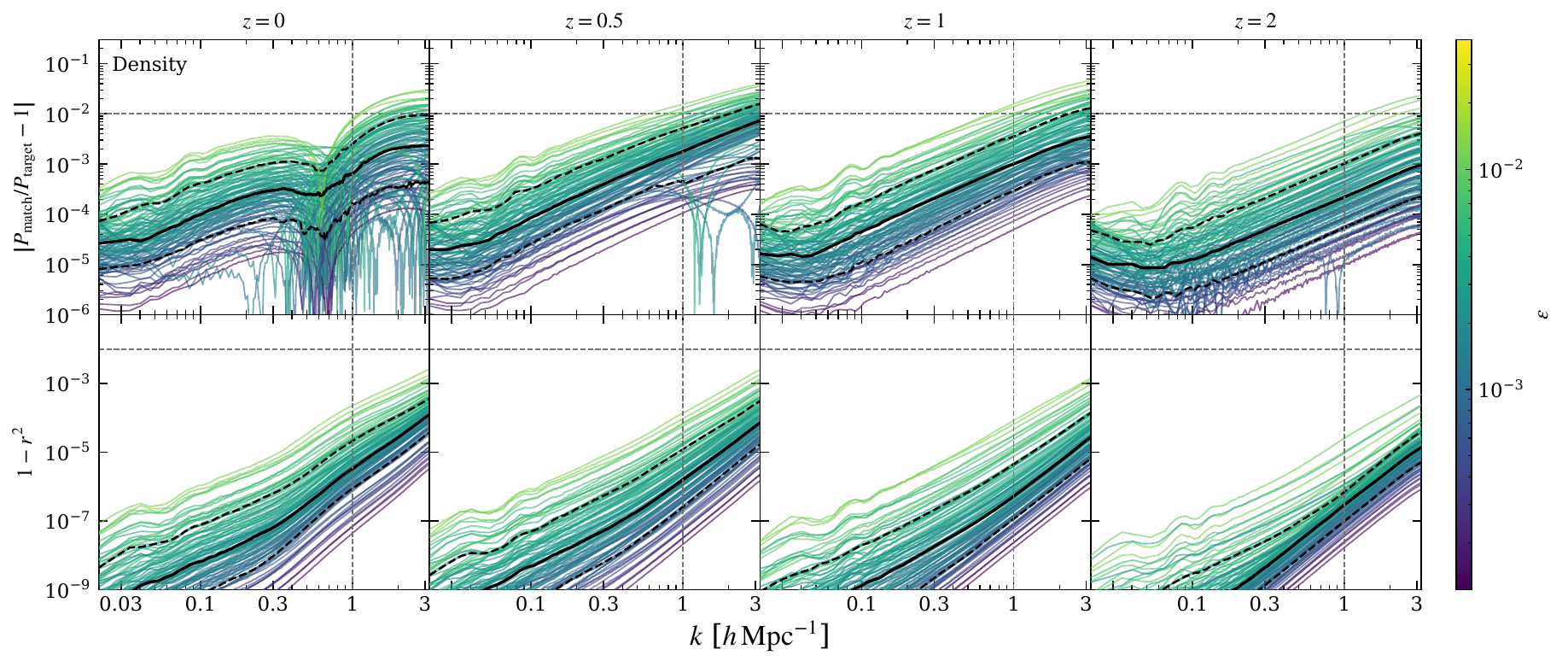}
    \caption{Power spectrum error $|\Delta P/P|$ (top) and stochasticity $1-r^2$ (bottom) of the density field for the 100 models with $w_a<0$ at $z=0$, 0.5, 1 and 2 (columns), coloured by $\epsilon$. Solid black lines show the median and dashed black lines the 16th and 84th percentiles across models. Grey dashed lines mark $10^{-2}$ and $k=1\,h\,{\rm Mpc}^{-1}$. The errors are strongly correlated with $\epsilon$, and for most models the power spectrum error is sub-percent and the stochasticity is $\sim10^{-4}$ or below up to the Nyquist frequency. The power spectra of target and match agree well.}
    \label{fig:ensemble}
\end{figure*}

The field-level difference can be characterised in terms of the stochasticity $1-r^2$ \citep{Hamaus2010}, where the cross-correlation coefficient $r$ is defined as
\begin{equation}
    r(k)\equiv P_{\rm cross}(k)\Big/\sqrt{P_{\rm target}(k)P_{\rm match}(k)},    
\end{equation}
with the cross power spectrum given as $\langle\delta_{\rm target}(\boldsymbol{k})\,\delta^*_{\rm match}(\boldsymbol{k}')\rangle=(2\pi)^3\delta_{\rm D}(\boldsymbol{k}-\boldsymbol{k}')\,P_{\rm cross}(k)$. The term $1-r^2$ quantifies the fraction of the variance in the target field that is unexplained by the matched field. It captures differences at the field level beyond the power spectrum, including differences in the phases of the modes, and is also used to validate field-level emulators. Beyond $k\approx1\,h\,{\rm Mpc}^{-1}$, baryonic effects become important \citep[e.g.][]{vanDaalen2011,Chisari2019}. We therefore focus on $k\lesssim1\,h\,{\rm Mpc}^{-1}$, but also show our results up to the Nyquist frequency of the mesh, $k_{\rm Ny}\approx3.2\,h\,{\rm Mpc}^{-1}$, for completeness.

Figure~\ref{fig:power} shows the power spectrum error, $\Delta P/P$, and the stochasticity of the density field, $1-r^2$. On linear scales, the power spectra agree by construction, since the growth factor is matched. On nonlinear scales, they agree to better than 0.25\% for $k\lesssim1\,h\,{\rm Mpc}^{-1}$ and to better than 0.9\% up to the Nyquist frequency, at all redshifts, with the greatest discrepancy observed at $z=0.5$. That $\Delta P/P$ at small scales initially increases and then decreases after $z=0.5$ is likely caused by the difference in the nonlinear growth rates between the target and matched models. The stochasticity grows with time but stays below $10^{-5}$ for $k\lesssim1\,h\,{\rm Mpc}^{-1}$, and reaches $10^{-4}$ only near the Nyquist frequency.

\begin{figure*}
    \centering
    \includegraphics[width=1\linewidth]{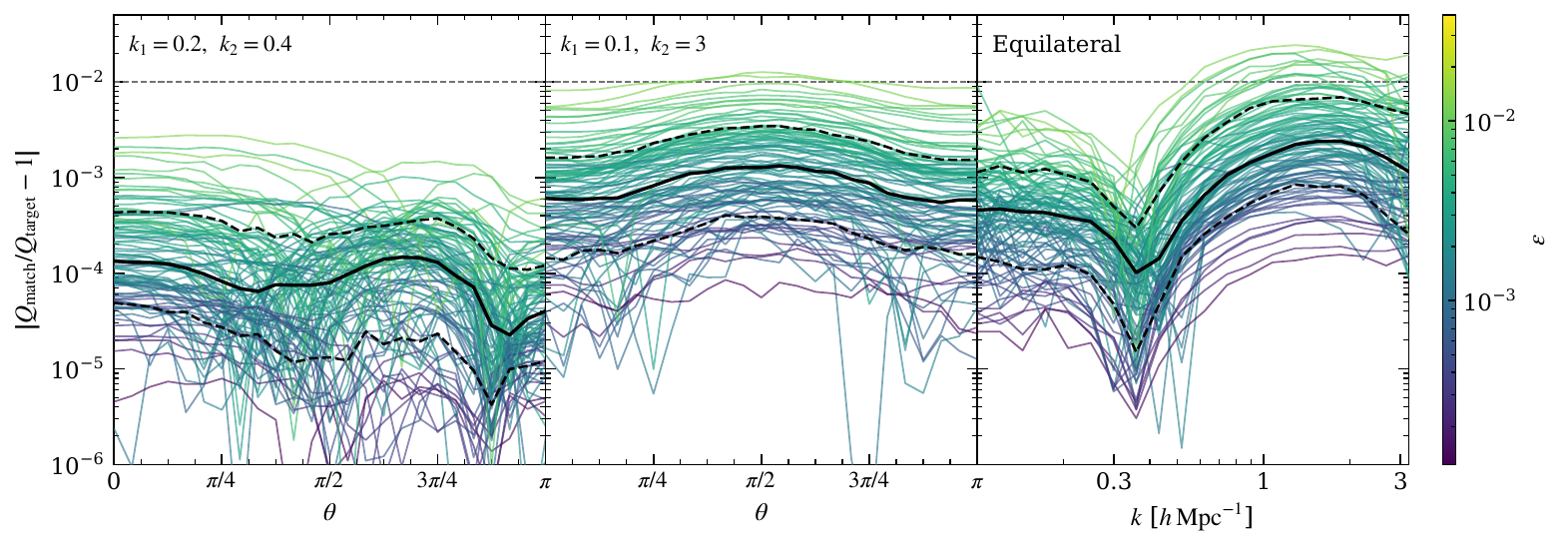}
    \caption{Reduced bispectrum error $|Q_{\rm match}/Q_{\rm target}-1|$ at $z=0$ for the models with $w_a<0$. Left and middle: triangles with fixed $(k_1,k_2)=(0.2,0.4)$ and $(0.1,3)\,h\,{\rm Mpc}^{-1}$, as a function of the angle $\theta$ between $\boldsymbol{k}_1$ and $\boldsymbol{k}_2$, with $k_3=|\boldsymbol{k}_1+\boldsymbol{k}_2|$. Right: equilateral triangles ($k_1=k_2=k_3=k$) as a function of $k$. Colours and lines as in Fig.~\ref{fig:ensemble}. The reduced bispectra of target and match agree well.}
    \label{fig:bispec}
\end{figure*}

\begin{figure*}
    \centering
    \includegraphics[width=1\linewidth]{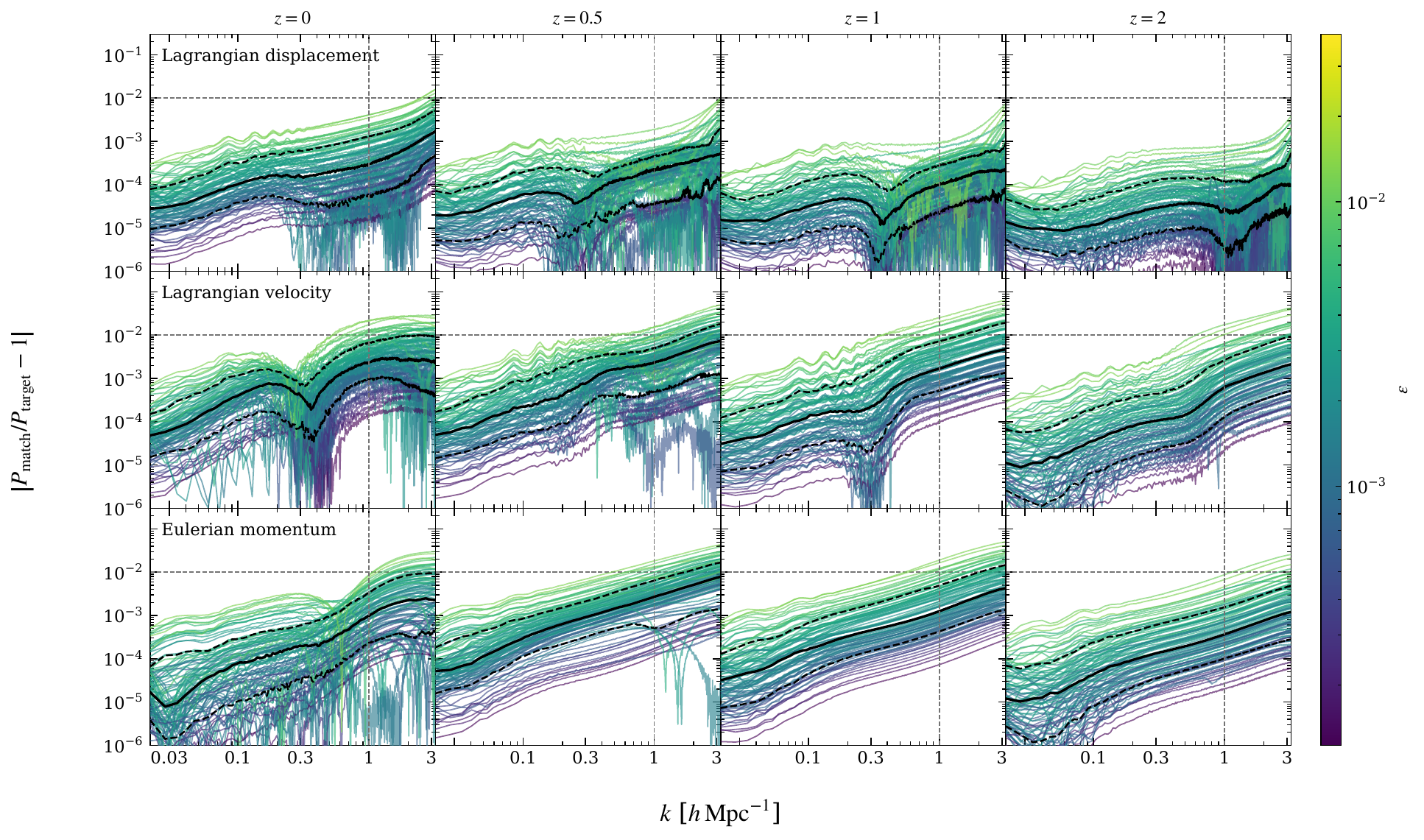}
    \caption{Power spectrum error $|\Delta P/P|$ of the Lagrangian displacement (top), Lagrangian velocity (middle) and Eulerian momentum (bottom), at $z=0$, 0.5, 1 and 2 (columns). Colours and lines as in Fig.~\ref{fig:ensemble}. The displacement is the most accurate field and the Lagrangian velocity the least accurate, but all three fields are sub-percent for $k\lesssim1\,h\,{\rm Mpc}^{-1}$ for most models.}
    \label{fig:fields}
\end{figure*}

Having examined the \ac{DESI}-like pair, we now turn to the full grid of 125 models (see Sect.~\ref{sec:simulations}), to assess how the accuracy of the matching depends on the target cosmology. We start by quantifying, in Fig.~\ref{fig:Jmap}, the error we make in approximating the expansion history by means of a constant $w$. The panels show the matching error $\epsilon$, defined in Eq.~\eqref{eq:J}, for $w_0\in[-1.4,-0.5]$, $w_a\in[-1.8,0.4]$ and three values of the matter density, $\Omega_{\rm m}\in\{0.1,0.3,0.5\}$. As a property of the background expansion history, we have a single $\epsilon$ for each model, and we expect the accuracy of the matched simulations to degrade as $\epsilon$ increases. We can see that $\epsilon$ is smallest in the $w_a=-0.15$ row, which is closest to a constant $w$. It increases with $w_0$ and $|w_a|$, and decreases with $\Omega_{\rm m}$, i.e.\ with a smaller dark energy fraction. These trends are expected: the matching is better where the target is closer to a constant $w$, and where dark energy is less important, i.e.\ for more negative $w_0$ or higher $\Omega_{\rm m}$.

We focus on the models with $w_a<0$ (boxed in Fig.~\ref{fig:Jmap}). Within this range, $\epsilon$ is largest in the lower-right corner, and the maximum is $\epsilon=0.0124$ at $(\Omega_{\rm m},w_0,w_a)=(0.1,-0.5,-1.8)$. When the models with positive $w_a$ are included, we find a much larger $\epsilon$ in the upper-right corner ($w_a=+0.4$, large $w_0$). In that region $w\rightarrow w_0+w_a$ is far above $-1$ at early times, so dark energy is not negligible at high redshift, which a constant $w$ is unable to reproduce. These models lie outside the \ac{DESI} contours \citep{DESI2025}, and we use them only as a stress test (Sect.~\ref{sec:extreme}). The \ac{DESI} DR2 constraint (\ac{DESI}+\ac{CMB}+DESY5, flat $w_0w_a$CDM, 68\% and 95\% contours), marginalised over $\Omega_{\rm m}$, is drawn on the $\Omega_{\rm m}=0.3$ panel, the closest to the \ac{DESI} value $\Omega_{\rm m}=0.3191\pm0.0056$, and it covers cells with $\epsilon\approx2$--$4\times10^{-3}$, where the matching is excellent.

Figure~\ref{fig:ensemble} shows the power spectrum error, $|\Delta P/P|$, and the stochasticity of the density field, $1-r^2$, for the 100 simulation pairs with $w_a<0$, at redshifts $z=0$, 0.5, 1 and 2, coloured by the matching error $\epsilon$. At all shown redshifts, the median error $|\Delta P/P|$ is below $0.2\%$ at all scales $k\lesssim 1\,h\,{\rm Mpc}^{-1}$ and below $0.8\%$ all the way up to the Nyquist frequency $k\lesssim 3\,h\,{\rm Mpc}^{-1}$. The median follows the behaviour of the \ac{DESI}-like pair: it increases from $z=3$ to a peak at $z=0.5$ and then decreases towards $z=0$, where a sign change near $k\approx0.6\,h\,{\rm Mpc}^{-1}$ appears as a dip. 
All models except six show sub-percent error in the power spectrum at all scales $k\lesssim 1\,h\,{\rm Mpc}^{-1}$ and at all redshifts, with the greatest discrepancy again appearing at $z=0.5$. For $k\gtrsim 1\,h\,{\rm Mpc}^{-1}$, where baryons nevertheless become important and the reliability of dark-matter-only simulations decreases, the maximum error stays within $5\%$, with the largest deviation shown at $z=1$.

The median stochasticity follows the same behaviour as in the \ac{DESI}-like pair, staying below $10^{-5}$ for $k\lesssim1\,h\,{\rm Mpc}^{-1}$ and reaching $10^{-4}$ only near the Nyquist frequency; it is also largest at $z=0$. For all models considered, the stochasticity remains sub-percent up to the Nyquist frequency. The errors are strongly correlated with $\epsilon$: all models below the 90th percentile of $\epsilon$, $\epsilon<6\times10^{-3}$, have a power spectrum error below 1\% and a stochasticity below $10^{-3}$ at all scales $k\lesssim1\,h\,{\rm Mpc}^{-1}$ and at all redshifts. All six exceptions have $\Omega_{\rm m}=0.1$ and lie in the top 10\% of $\epsilon$.

\begin{figure*}
    \centering
    \includegraphics[width=1\linewidth]{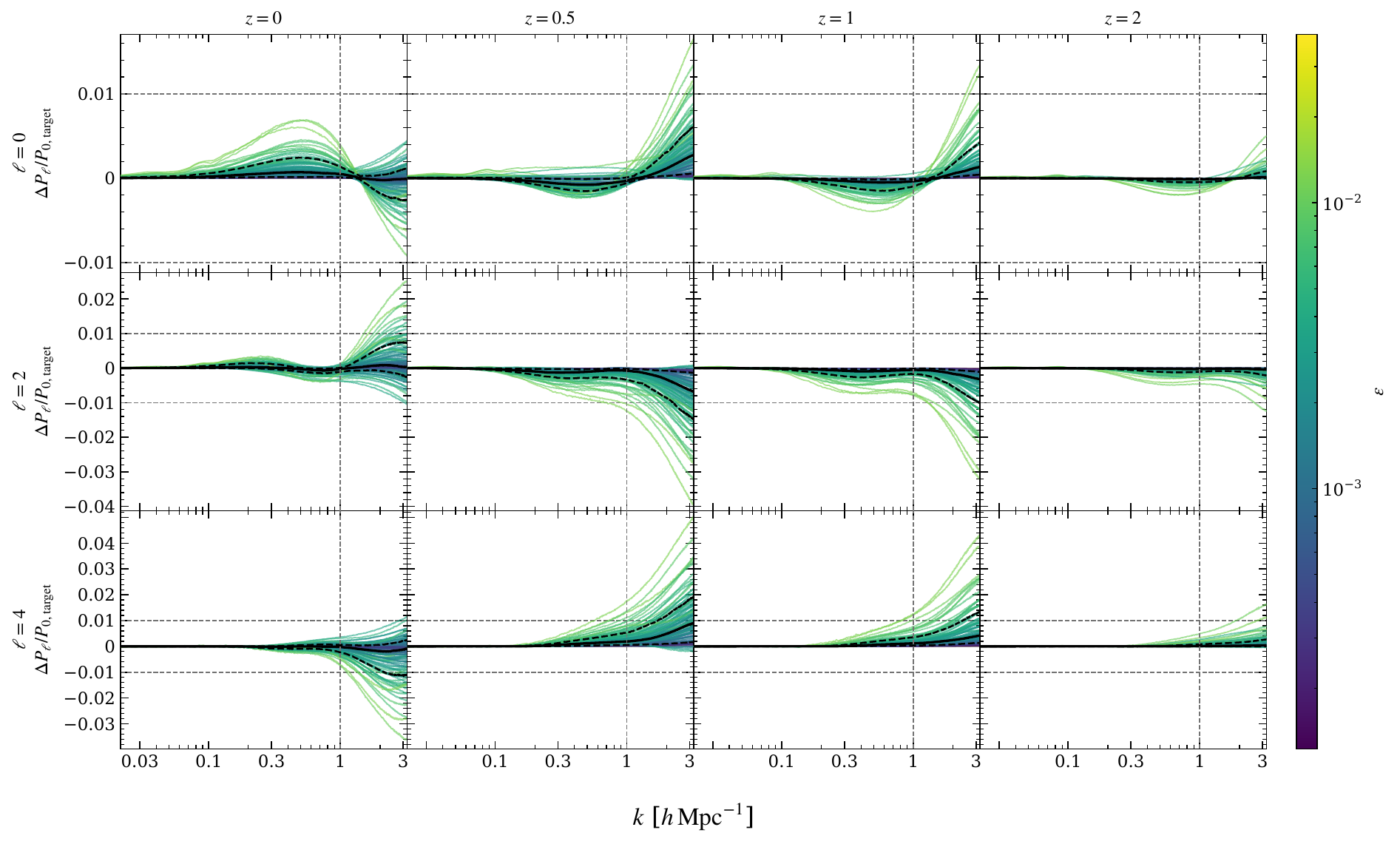}
    \caption{Residuals of the redshift-space monopole, quadrupole and hexadecapole (rows), normalised by the target monopole and averaged over three lines of sight, at $z=0$, 0.5, 1 and 2 (columns). Colours and lines as in Fig.~\ref{fig:ensemble}. The multipoles agree to within 1\% at $k\lesssim1\,h\,{\rm Mpc}^{-1}$ for nearly all models. Towards the Nyquist frequency, the residuals of the quadrupole and hexadecapole reach a few percent for the small number of models with the largest $\epsilon$.}
    \label{fig:rsd}
\end{figure*}

To test the non-Gaussianity of the field, we use the bispectrum, defined by $\langle\delta(\boldsymbol{k}_1)\delta(\boldsymbol{k}_2)\delta(\boldsymbol{k}_3)\rangle=(2\pi)^3\delta_{\rm D}(\boldsymbol{k}_1+\boldsymbol{k}_2+\boldsymbol{k}_3)\,B(k_1,k_2,k_3)$ \citep{Bernardeau2002,Scoccimarro2000}. In particular, we focus on the reduced bispectrum $Q\equiv B/[P(k_1)P(k_2)+P(k_2)P(k_3)+P(k_3)P(k_1)]$, which largely removes the dependence on the power spectrum amplitude and isolates the mode coupling. We consider three types of triangle configurations, which probe different aspects of mode coupling. Two of them fix $k_1$ and $k_2$ and vary the angle $\theta$ between them. The first, $k_2=2k_1=0.4\,h\,{\rm Mpc}^{-1}$, probes the quasi-linear regime. The second, $(k_1,k_2)=(0.1,3)\,h\,{\rm Mpc}^{-1}$, is a squeezed configuration, which couples a large-scale mode to two nonlinear modes and therefore tests how small-scale structure responds to its large-scale environment. Finally, the equilateral configuration, from quasi-linear scales up to just below $k_{\rm Ny}$, follows the mode coupling at fixed shape from quasi-linear to nonlinear scales. We measure the bispectrum at $z=0$ with \textsc{pylians3} \citep{VillaescusaNavarro2018} on the same $1024^3$ \ac{CIC} mesh, correcting for the \ac{CIC} window function. Shot noise is not subtracted, since it is nearly identical for target and match.

Figure~\ref{fig:bispec} shows the error of the reduced bispectrum, $|\Delta Q/Q|\equiv|Q_{\rm match}/Q_{\rm target}-1|$, at $z=0$. For the quasi-linear configuration (left panel), all models agree to better than 0.3\% at all angles, with a median below 0.02\%.
When the large-scale mode $k_1=0.1\,h\,{\rm Mpc}^{-1}$ is combined with $k_2=3\,h\,{\rm Mpc}^{-1}$ (middle panel), the residual depends only weakly on $\theta$ and is largest near $\theta\approx\pi/2$, with a median below 0.15\%. All models except two stay below 1\%, with a maximum of 1.3\%. For equilateral triangles (right panel), the median error is below 0.15\% for $k<1\,h\,{\rm Mpc}^{-1}$ and below 0.25\% at all $k$, with the largest values at $k\approx1.5$--$2\,h\,{\rm Mpc}^{-1}$. All models except ten stay below 1\%, with a maximum of 2.4\%. The residuals of $Q$ are strongly correlated with $\epsilon$, similar to those of the power spectrum.

\begin{figure*}
    \centering
    \includegraphics[width=1\linewidth]{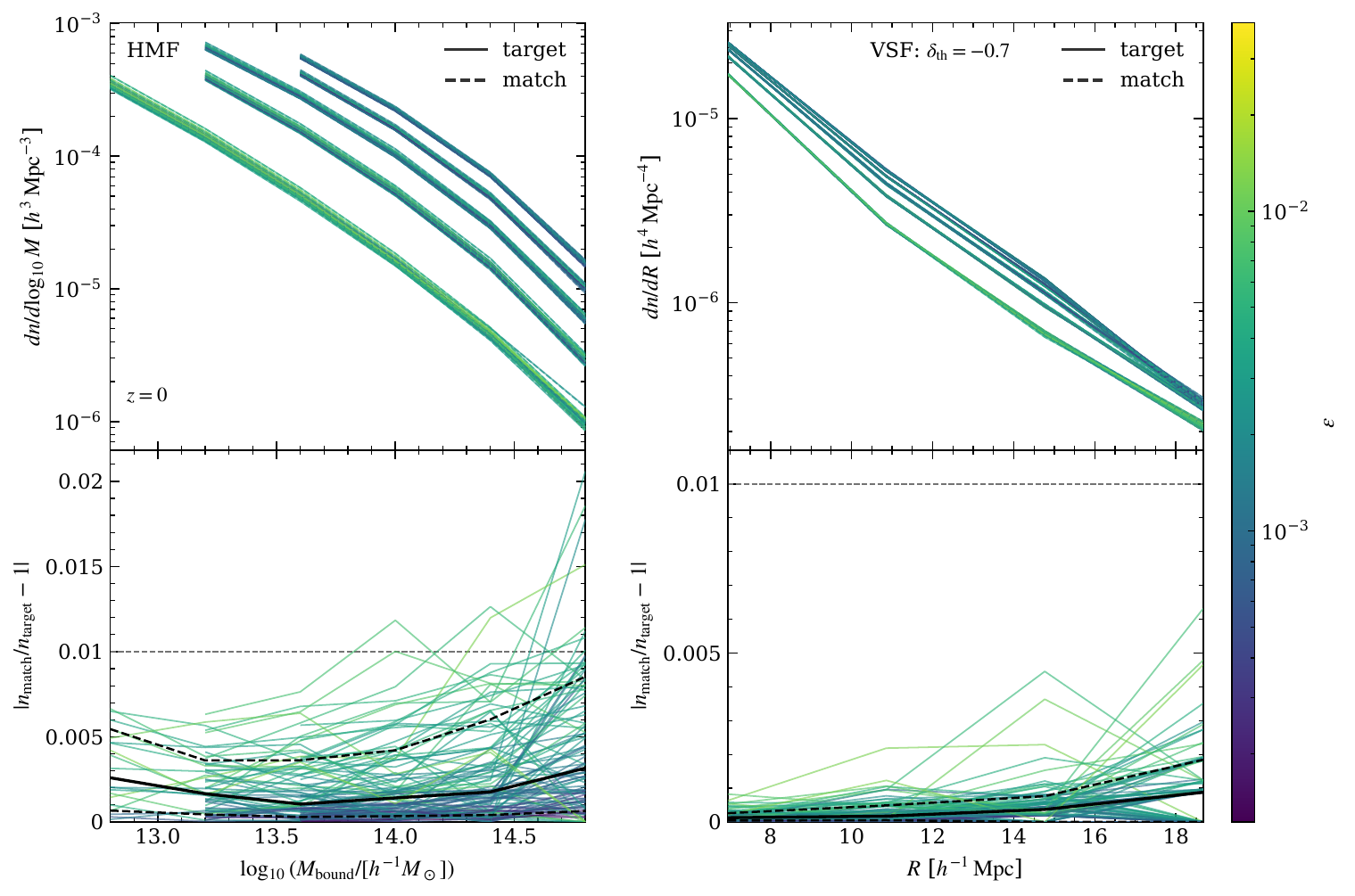}
    \caption{Halo mass function (HMF, left) and void size function (VSF, right) at $z=0$, where voids are spheres with mean overdensity below $\delta_{\rm th}-0.7$. Top: target (solid) and match (dashed). Bottom: $|n_{\rm match}/n_{\rm target}-1|$ for bins with at least 100 objects in both members, where zero means equal counts. Haloes are binned in 0.4 dex in mass up to $10^{15}\,h^{-1}M_\odot$ and voids in $3.9\,h^{-1}{\rm Mpc}$ in radius up to $R=20.6\,h^{-1}{\rm Mpc}$. Colours and lines as in Fig.~\ref{fig:ensemble}. Grey dashed lines mark 1\%.}
    \label{fig:hmf}
\end{figure*}

\begin{figure*}
    \centering
    \includegraphics[width=1\linewidth]{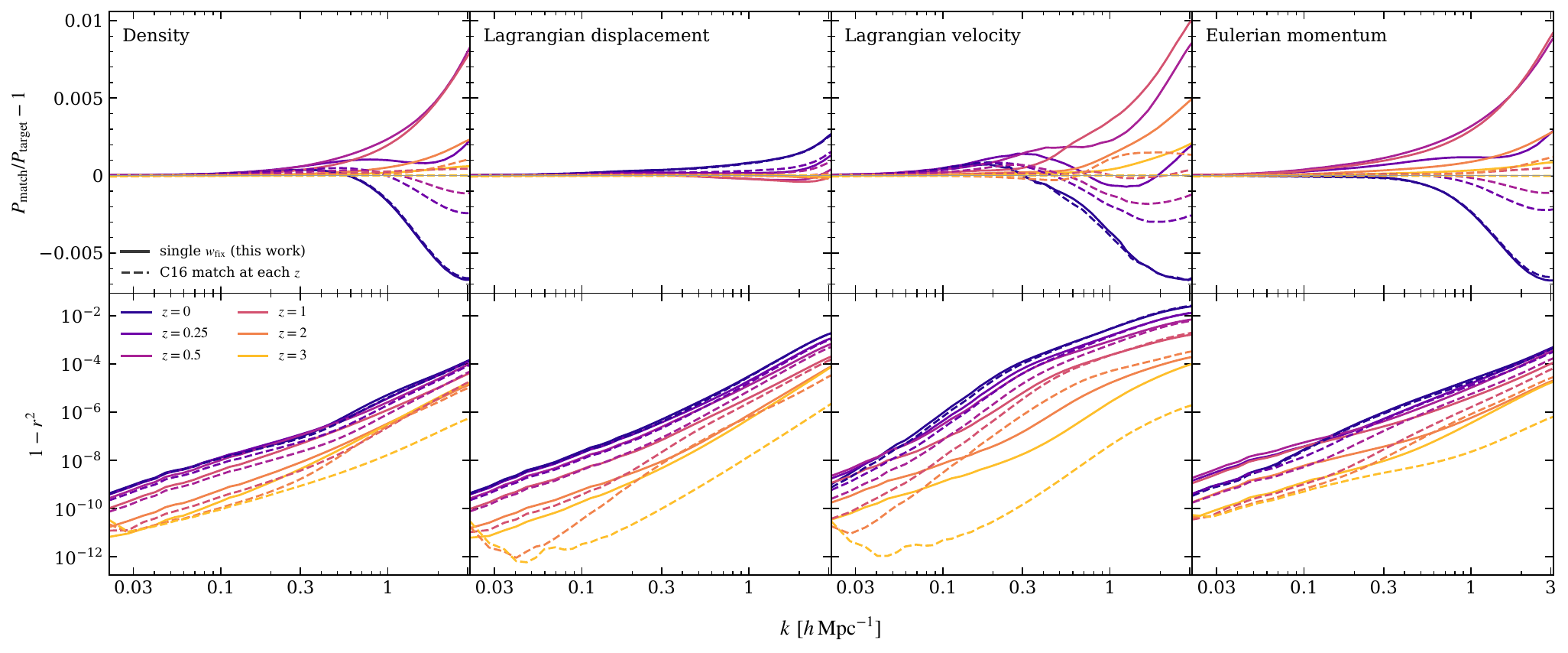}
    \caption{Comparison between our (solid) and \citetalias{Casarini2016}'s redshift-dependent (dashed) matching for the DESI-like target $(\Omega_{\rm m},w_0,w_a)=(0.3191,-0.752,-0.86)$. Our matching obtains all snapshots from a single constant-$w$ simulation, whereas \citetalias{Casarini2016} requires a separate constant-$w$ simulation, with its own $w$ and $\sigma_8$, for each redshift. At $k\lesssim1\,h\,{\rm Mpc}^{-1}$, the two approaches are close, with $|\Delta P/P|<0.4\%$ for all fields. Towards the Nyquist frequency, the density error of our matching grows to about 0.9\% at $z=0.5$ and 1, compared with at most 0.12\% for \citetalias{Casarini2016}. Our matching thus trades a small loss in accuracy for a large gain in efficiency. Top: power spectrum error $P_{\rm match}/P_{\rm target}-1$. Bottom: stochasticity $1-r^2$. Columns show the density, Lagrangian displacement, Lagrangian velocity and Eulerian momentum. The velocities of both approaches are rescaled with Eq.~\eqref{eq:velcorr}. Colours denote $z$ as indicated by the legend.}
    \label{fig:se}
\end{figure*}

\subsection{Lagrangian, momentum and redshift-space fields}
\label{sec:fields}

In addition to the density, we examine the Lagrangian displacement and velocity and the Eulerian momentum field (see Sect.~\ref{sec:simulations}), which are vector fields. For a vector field $\boldsymbol{u}$, which is $\boldsymbol{\Psi}(\boldsymbol{q})$, $\boldsymbol{v}(\boldsymbol{q})$ or $\boldsymbol{p}(\boldsymbol{x})$, we define $P_{\boldsymbol{u}}(k)=\sum_{i=1}^{3}P_{u_iu_i}(k)$, and likewise for the cross power. The cross-correlation coefficient is thus
\begin{equation}\label{eq:rvec}
r(k)=\frac{\sum_{i}P_{u_i^{\rm target}u_i^{\rm match}}(k)}{\left[\sum_{i}P_{u_i^{\rm target}u_i^{\rm target}}(k)\,\sum_{i}P_{u_i^{\rm match}u_i^{\rm match}}(k)\right]^{1/2}},
\end{equation}
which treats the three Cartesian components as a single field. The velocities and momentum of the match are rescaled as in Sect.~\ref{sec:matching} (see Eq.~\eqref{eq:velcorr}); we also discuss the implications of rescaling further in Appendix~\ref{app:velcor}.

Figure~\ref{fig:fields} shows the power spectrum error $|\Delta P/P|$ of the Lagrangian displacement, Lagrangian velocity and Eulerian momentum and its dependence on redshift. The displacement is the most accurate of the three fields. Its median error is below 0.04\% for $k\lesssim1\,h\,{\rm Mpc}^{-1}$ and below 0.2\% up to the Nyquist frequency. All models stay below 0.5\% for $k\lesssim1\,h\,{\rm Mpc}^{-1}$ and below 1.6\% up to the Nyquist frequency. The displacement error grows with time and is largest at $z=0$. For both the Lagrangian velocity and the momentum, the median error is below 0.3\% for $k\lesssim1\,h\,{\rm Mpc}^{-1}$ and below 0.8\% up to the Nyquist frequency. For $k\lesssim1\,h\,{\rm Mpc}^{-1}$, at most eleven models exceed 1\% in either field, and these are among the models with the largest $\epsilon$. The Lagrangian velocity is the least accurate field, with errors up to 2.6\% for $k\lesssim1\,h\,{\rm Mpc}^{-1}$ and 6.4\% up to the Nyquist frequency. We will revisit this point in Appendix \ref{app:codes}.

\begin{figure*}
    \centering
    \includegraphics[width=1\linewidth]{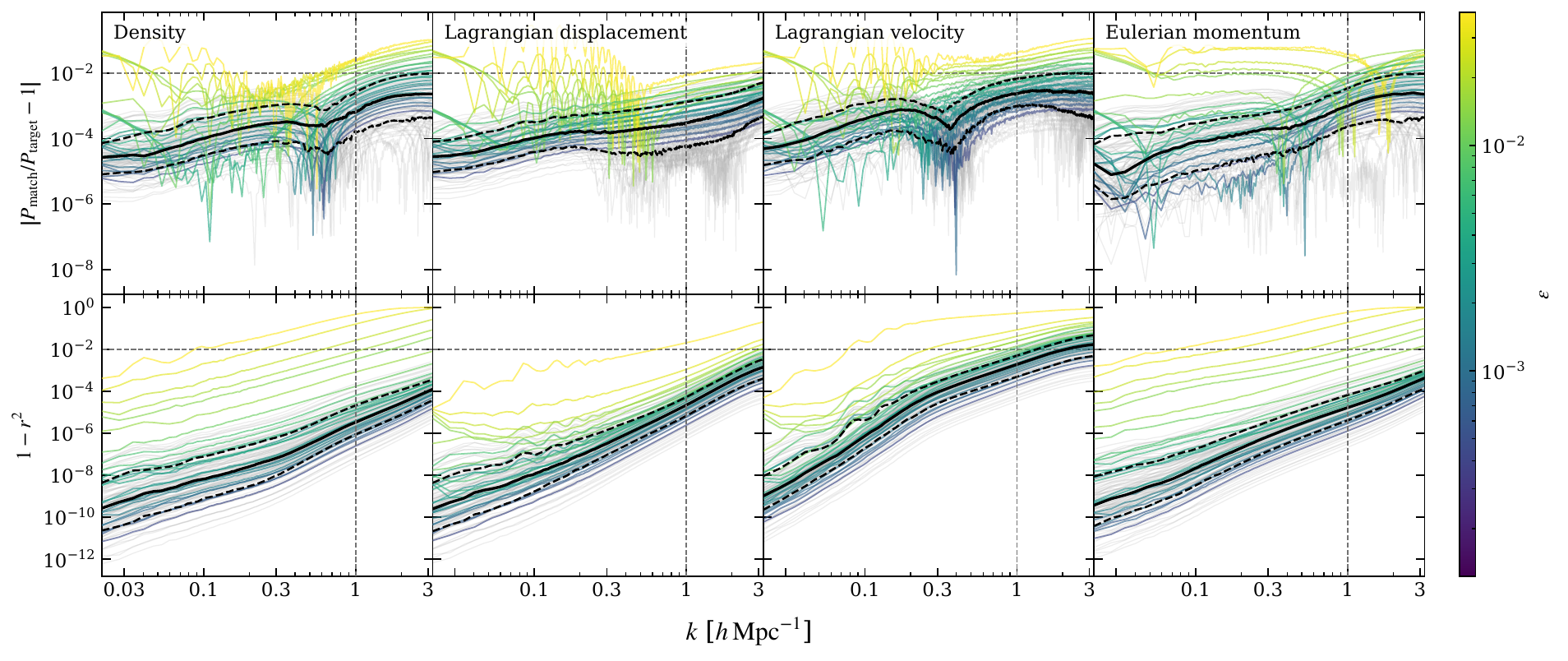}
    \caption{Power spectrum error $|\Delta P/P|$ (top) and stochasticity $1-r^2$ (bottom) at $z=0$ of the density, Lagrangian displacement, Lagrangian velocity and Eulerian momentum for the 25 simulation pairs with $w_a=+0.4$, coloured by $\epsilon$. These models have non-negligible dark energy density at early times, making them difficult to reproduce with a fixed equation of state. Grey lines show the 100 simulation pairs with $w_a<0$, and the black solid and dashed lines their median and 16th and 84th percentiles. Grey dashed lines mark $10^{-2}$. In this stress test, the models with the largest $\epsilon$ also have the largest spectrum and field-level errors.}
    \label{fig:positive_wa}
\end{figure*}

Observations measure positions in redshift space, so we also compare the redshift-space density fields. We construct them in the plane-parallel approximation, taking each of the three box axes in turn as the line of sight, average the multipoles over the three directions, and compare target and match. The multipoles are the Legendre expansion of the redshift-space power spectrum $P(k,\mu)$, where $\mu$ is the cosine of the angle to the line of sight \citep{Kaiser1987,Hamilton1998}
\begin{equation}
    P_\ell(k) = \frac{2\ell+1}{2}\int_{-1}^{1}P(k,\mu)\,\mathcal{L}_\ell(\mu)\,\mathrm{d}\mu.
\end{equation}
Here $\mathcal{L}_\ell$ is the Legendre polynomial of order $\ell$, and we measure the monopole ($\ell=0$), quadrupole ($\ell=2$) and hexadecapole ($\ell=4$). We present the residuals of all multipoles normalised by the target monopole, $\Delta P_\ell/P_0\equiv(P_{\ell,{\rm match}}-P_{\ell,{\rm target}})/P_{0,{\rm target}}$. The velocities used to construct the redshift-space positions are rescaled following Eq.~\eqref{eq:velcorr} (see also Appendix~\ref{app:velcor}).

Figure~\ref{fig:rsd} shows the residuals of the redshift-space monopole, quadrupole and hexadecapole, $\Delta P_\ell/P_0$. For $k\lesssim1\,h\,{\rm Mpc}^{-1}$, the median $|\Delta P_\ell/P_0|$ is below 0.2\% for all three multipoles. The residuals are within 1\% for all models, except for the six models that also exceed 1\% in the real-space power spectrum (Sect.~\ref{sec:density}), for which the hexadecapole reaches 1.8\%. Up to the Nyquist frequency, the median remains below 1\%. The largest residuals are in the hexadecapole at $z=0.5$, with a maximum of about 5\%. The matching is therefore as accurate in redshift space as in real space.

\subsection{Halo and void statistics}
\label{sec:haloes}

We compare the abundance of haloes at $z=0$ (Fig.~\ref{fig:hmf}), following the procedure described in Sect.~\ref{sec:simulations}. To exclude poorly resolved haloes and sparsely populated bins, we restrict the comparison to masses from the 100-particle mass of each cosmology up to $10^{15}\,h^{-1}M_\odot$, where target and match contain at least 100 haloes per bin. The median $|n_{\rm match}/n_{\rm target}-1|$ is $\lesssim0.3\%$ at all masses and largest at the highest masses, $10^{14.6}$--$10^{15}\,h^{-1}M_\odot$, where haloes are rare, and the 84th percentile stays below 1\%. Since target and match share the same initial phases, the Poisson noise largely cancels in the ratio, which allows a sub-percent comparison even where haloes are rare.

We also compare the void abundance at $z=0$ (Fig.~\ref{fig:hmf}). We use the void finding procedure described in Sect.~\ref{sec:simulations}. The comparison covers $R=5$--$20.6\,h^{-1}{\rm Mpc}$, where the target and match contain at least 100 voids per bin. The median $|n_{\rm match}/n_{\rm target}-1|$ increases with radius, from 0.01\% at $R\approx5$--$9\,h^{-1}{\rm Mpc}$ to 0.09\% at $R\approx17$--$21\,h^{-1}{\rm Mpc}$, and the 84th percentile stays below 0.2\%. The fraction of models in which target and match contain exactly the same number of voids increases from 5\% at the smallest radii to 48\% at the largest. Voids are reproduced more precisely than haloes, as expected for large, quasi-linear structures. For both statistics, the largest residuals belong to the models with the largest $\epsilon$. 
\subsection{Comparison with the redshift-dependent matching of \citetalias{Casarini2016}}
\label{sec:se}

As mentioned in Sect.~\ref{sec:matching}, our matching trades a little accuracy for efficiency: a single constant-$w$ model serves all redshifts, so that one $N$-body simulation provides all the snapshots of the target. The matching of \citetalias{Casarini2016}, instead, treats each output redshift separately, with a constant-$w$ model that has the same comoving distance to the last-scattering surface and the same $\sigma_8(z)$ as the target. This reproduces the nonlinear matter power spectrum at each redshift to better than 1\% up to $k=3\,h\,{\rm Mpc}^{-1}$, but requires a separate $N$-body simulation, with its own $w$ and $\sigma_8$, for every snapshot. For the \ac{DESI}-like target, the \citetalias{Casarini2016} models range from $w=-0.9759$ at $z=0$ to $w=-1.1872$ at $z=3$. We run these simulations with \textsc{mg-glam}, using the same initial phases and the same velocity rescaling (Eq.~\eqref{eq:velcorr}) as for our match, and compare the two approaches in Fig.~\ref{fig:se}, to quantify the accuracy we give up by using a single simulation.

At $z=0$, the \citetalias{Casarini2016} model, with $w=-0.9759$, is very close to our match, $w_{\rm fix}=-0.9748$, and the two approaches give nearly identical results. For all four fields, their power spectrum errors differ by less than 0.05\% for $k\lesssim1\,h\,{\rm Mpc}^{-1}$, 
and the density stochasticity stays below $5.3\times10^{-6}$ for our match and $4.0\times10^{-6}$ for \citetalias{Casarini2016} in the same range. At $z>0$, \citetalias{Casarini2016} is generally more accurate, in particular on small scales. The difference is largest at $z=0.5$ and 1, 
where the density $|\Delta P/P|$ up to near the Nyquist frequency reaches 0.86\% and 0.81\% for our match, compared with 0.12\% and 0.045\% for \citetalias{Casarini2016}. These are the epochs at which the expansion history of our match departs most from that of the target, with $|\Delta\ln a|$ up to 0.7\% near $z\approx0.7$ (Fig.~\ref{fig:matching}, left panel). For $k\lesssim1\,h\,{\rm Mpc}^{-1}$, the density $|\Delta P/P|$ at $z>0$ stays below 0.24\% for our match and 0.05\% for \citetalias{Casarini2016}. The momentum and the Lagrangian velocity behave in the same way, while the displacement errors of the two approaches are similar, $\lesssim0.04\%$ for $k\lesssim1\,h\,{\rm Mpc}^{-1}$. The density stochasticity of \citetalias{Casarini2016} is also lower at $z>0$, although both remain below $10^{-5}$ for $k\lesssim1\,h\,{\rm Mpc}^{-1}$. 

The two approaches therefore trade accuracy against efficiency. \citetalias{Casarini2016} is slightly more accurate at $z>0$, in particular on small scales, but needs a separate simulation, with a different cosmology, for every output redshift, six in this comparison. Our matching uses a single simulation of one constant-$w$ cosmology for all redshifts, and its outputs form one continuous evolution, which is suitable for building light cones. For a small loss in accuracy, with errors of a few tenths of a percent for $k\lesssim1\,h\,{\rm Mpc}^{-1}$ for the \ac{DESI}-like target, we thus gain efficiency, and a constant-$w$ simulation suite or field-level emulator can reproduce a \ac{CPL} target at all redshifts with a single input cosmology. We also note that, for $k\gtrsim1\,h\,{\rm Mpc}^{-1}$, we are moving to the one-halo regime, where sub-percent agreement is difficult to achieve between different simulation codes, particularly at field level \citep{Bayer2025}. 

\subsection{Extreme cases}
\label{sec:extreme}

The 25 simulation pairs with $w_a=+0.4$ test the matching outside the parameter range favoured by \ac{DESI}. They are the more extreme cases, with matching errors up to $\epsilon=0.039$ for $(\Omega_{\rm m},w_0,w_a)=(0.1,-0.5,0.4)$, three times the maximum for the models with $w_a<0$. Figure~\ref{fig:positive_wa} shows the results at $z=0$, where the models with the largest $\epsilon$ also have the largest spectrum and field-level errors. 
For $k\lesssim1\,h\,{\rm Mpc}^{-1}$, six of the 25 pairs have a density power spectrum error above 1\%, and five of them also have a stochasticity above 0.1\%, up to 46\%. These six pairs are the five with $(w_0,w_a)=(-0.5,0.4)$, one for each $\Omega_{\rm m}$, and $(0.1,-0.725,0.4)$, all in the upper-right corner of the $\epsilon$ map (Fig.~\ref{fig:Jmap}). The redshift-space multipoles, the reduced bispectrum and the halo mass function of these models also deviate by more than 1\%.

We note the worse matching in these more extreme models is not an indication of failure by the general methodology developed here, but simply the consequence of the fact that a constant-$w_{\rm fix}$ expansion history fundamentally cannot well match the expansion history of the target. The failing models have a non-negligible dark energy density at early times. For the model with the largest $\epsilon$, $(0.1,-0.5,0.4)$, the equation of state approaches $w_0+w_a=-0.1$ at early times, and dark energy makes up 41\% of the energy density at $z_{\rm init}=100$. For the five models with $(w_0,w_a)=(-0.5,0.4)$, this fraction increases from 7\% for $\Omega_{\rm m}=0.5$ to 41\% for $\Omega_{\rm m}=0.1$. For $(0.1,-0.725,0.4)$, it is 3\%, compared with less than 0.2\% for every model with $w_a<0$. A constant $w$ cannot follow a dark energy component that is important both at early and at late times. The matches of these models have $w_{\rm fix}=-0.26$ to $-0.28$ and a dark energy fraction of 2--19\% at $z=100$, so the expansion histories, and hence the particle trajectories, differ already when structures start to form. Early dark energy also changes nonlinear structure formation beyond the linear growth \citep{Francis2009,Grossi2009}, and fitting functions for the nonlinear power spectrum show their largest errors in the same corner of parameter space, where $w_0+w_a$ approaches zero \citep{Mead2021}.

These cases therefore provide a stress test of our matching method with the additional constraint of $w=w_{\rm fix}$ in the matched cosmology. A single constant $w$ matches the expansion history only on average. This works well for the models with $w_a<0$. For the models with $w_a>0$ and a non-negligible early dark energy fraction, the matching could still work, but requires more sophisticated matched models than $w_{\rm fix}$ and so is less useful in this particular context. This limitation is of little practical concern at present, as the region of parameter space favoured by current data lies at $w_a<0$ \citep{DESI2025}. 
\section{Conclusions}\label{sec:conclusions}

We have presented a method to reproduce the full output of an $N$-body simulation with \ac{CPL} dark energy at multiple redshifts, using a single simulation of a constant-$w$ cosmology. The mapping between the two is obtained by matching their expansion histories in superconformal time.

We assessed the accuracy of the method with \textsc{mg-glam} simulations of a \ac{DESI}-like pair, with the target at the \ac{DESI}+\ac{CMB}+DESY5 posterior mean, and of 100 simulation pairs spanning $(\Omega_{\rm m},w_0,w_a)$ with $w_a<0$, which include the region favoured by \ac{DESI}. Our main results are as follows.

\begin{itemize}
\item \textit{\ac{DESI}-like pair.} For a target at the posterior mean of the \ac{DESI} constraints, the expansion histories of the target and the match differ by less than 1\% at all times. A single matched simulation reproduces the target at all output redshifts, $0\le z\le3$. The power spectra of the matter density, the Lagrangian displacement and velocity, and the Eulerian momentum agree to sub-percent precision up to near the Nyquist frequency of our mesh. At the field level, the density fields of the target and the match are almost perfectly correlated, and their one-point distributions agree to sub-percent precision.

\item \textit{Models with $w_a<0$.} For the 100 simulation pairs with $w_a<0$, the median error of the matter power spectrum is sub-percent up to near the Nyquist frequency, and the stochasticity of the density field is sub-percent. For $k\lesssim1\,h\,{\rm Mpc}^{-1}$, the power spectra of all fields and the redshift-space multipoles agree to sub-percent precision, except for the models with the largest matching errors. The reduced bispectrum and the abundances of haloes and voids also agree to sub-percent precision for most models. The errors of all statistics are strongly correlated with the matching error $\epsilon$, which depends only on the background. The models near the \ac{DESI} constraints have small $\epsilon$ and sub-percent errors.

\item \textit{Comparison with \citetalias{Casarini2016}.} For the \ac{DESI}-like pair, the redshift-dependent matching of \citetalias{Casarini2016} gives nearly identical results at $z=0$ and is more accurate on small scales at $z>0$, at the expense of requiring a separate simulation for every output redshift. Our matching needs a single simulation for all redshifts, with sub-percent errors for $k\lesssim1\,h\,{\rm Mpc}^{-1}$ in all fields. It therefore trades a small loss in accuracy for a large gain in efficiency.
\end{itemize}

Our method allows existing constant-$w$ simulation suites and field-level emulators to be used for CPL cosmologies without retraining, with mapping errors below the percent-level accuracy of most current field-level emulators. It is therefore naturally suited to building future emulators, which could be trained on a reduced set of models and extended to other models by matching.

Two points mentioned briefly above merit a more detailed discussion. First, when implementing the matching, we required that the matched and target cosmologies have the same $\sigma_8$ at $z=0$, which means that they start from initial conditions with slightly different amplitude $A_{\rm s}$ (but the same $n_{\rm s}$ and random phases). This is a choice: the alternative is to use the same initial conditions (those of the target cosmology) for both simulations. That would have been a natural choice, because the matched simulation is only a surrogate, which does not necessarily carry physical content. While we have not tested this approach here, we expect it to give better agreement at higher $z$ and worse at low $z$, with the overall accuracy again depending on $\epsilon$. Our choice is based on a practical consideration: it allows us to repurpose existing constant-$w$ simulations to build emulators for \ac{CPL}, without having to run new simulations.

Second, while the general matching method proposed here is exact, in the first application we have adopted constant-$w$ for the matched models, for its simplicity. This is because \ac{CPL} itself is not a complicated model, and using a more sophisticated matched model would lose the advantage of the method. Indeed, the matching method here is strikingly simple: the matched models have identical cosmological parameters as the targets, such as $h$, $\Omega_{\rm m}$, $\Omega_{\rm b}$, $\sigma_8$ and $n_{\rm s}$. For more complex models, such as other forms of time-dependent dark energy \ac{EoS}, coupled scalar fields (with no screening), interacting dark energy and the usual $\mu$--$\Sigma$ parameterisation of modified gravity \citep[e.g.][]{Pogosian2016}, this method will still apply, as we will demonstrate in a forthcoming work, but the matched model will have different cosmological parameters, and $\tau$ might need to be shifted or rescaled to achieve an accurate matching. This, however, is not necessarily a disadvantage, for a few reasons. (1) the additional flexibility can allow the matching to be more accurate, i.e.\ $\epsilon$ smaller, which, as we have found, is the deciding factor for the accuracy of the method; (2) a single suite of surrogate simulations can then be used to build field-level emulators for a variety of \ac{DE} and modified gravity models, making the method even more universal and efficient; (3) if a new suite of surrogate simulations is run, we could include the amplitude of the primordial power spectrum as an additional emulation parameter. This would turn the initial conditions into an extra dimension of the parameter space, rather than tying them to a matched model, and allow us to use the same initial conditions for the target and matched models. Together with a more accurate matching, this may remove the need for the growth factor matching and the velocity rescaling.

Finally, we stress that the matching works because the background enters the $N$-body equations through a single function of time, $a(\tau)$. The $N$-body equations, Eq.~\eqref{eq:eom}, are then entirely scale-free. Models with more freedom, such as those with massive neutrinos or scale-dependent modified gravity, cannot be described by $a(\tau)$ alone, so the method here will not apply straightforwardly. We leave adaptations of the method to these models for future work. 


\textit{Note added.}
While completing this manuscript, we became aware of the related work of Demidov et al.\ (in prep.), which also examined reproducing dynamical dark energy by means of a fixed-$w$ cosmology at the field level. They adopt the matching of \citetalias{Casarini2016}, and test it at $z=0$ for a \ac{DESI}-like \ac{CPL} model and two $w(z)$ cosmologies beyond \ac{CPL}. Our work differs in that the matching follows from an exact equivalence of the $N$-body equations in different models when expressed in superconformal time, which also identifies the origin of the error as the mismatch in $a(\tau)$, and in that a single matched simulation gives all redshifts. With our method, the mismatch is quantified by a single number, $\epsilon$, depending only on the background, with which the errors of all the summary statistics we tested are strongly correlated. These two studies are complementary. Demidov et al.\ explore the range of \ac{DE} models and tracer types: at $z=0$, they consider beyond-\ac{CPL} \ac{EoS} and extend the comparison to individual haloes and to the redshift-space clustering of mock galaxies. We explore the range of cosmologies and redshifts: for \ac{CPL}, we cover a grid in $(\Omega_{\rm m},w_0,w_a)$ at $0\le z\le3$, and also test the Lagrangian displacement and velocity fields and the dependence on the $N$-body code. Both methods work for general non-clustering \ac{DE} models not limited to \ac{CPL}.

\section*{Acknowledgements}

We thank Jie Wang and Sownak Bose for helpful discussions, and Anatoly Klypin and Paco Prada for sharing the \textsc{glam} code, which \textsc{mg-glam} is based on.

YM is supported by a China Scholarship Council studentship (No. 202504910166). YM, DS, C-ZR and BL are supported by the European Research Council (ERC) Advanced Grant ``UNCA'' (UKRI Frontiers Research Guarantee No.~EP/Z533877/1). KK is supported by UK STFC grant number ST/B001175/1.

This work used the DiRAC@Durham facility managed by the Institute for Computational Cosmology on behalf of the STFC DiRAC HPC Facility (www.dirac.ac.uk). The equipment was funded by BEIS capital funding via STFC capital grants ST/K00042X/1, ST/P002293/1, ST/R002371/1 and ST/S002502/1, Durham University and STFC operations grant ST/R000832/1. DiRAC is part of the National e-Infrastructure.

For the purpose of open access, we have applied a Creative Commons Attribution (CC BY) licence to any Author Accepted Manuscript version arising.

\section*{Data Availability}

The supporting research data are available upon reasonable request from the corresponding author.



\bibliographystyle{mnras}
\bibliography{references}



\appendix
\section{A comparison of $N$-body codes}
\label{app:codes}

\begin{figure*}
    \centering
    \includegraphics[width=1\linewidth]{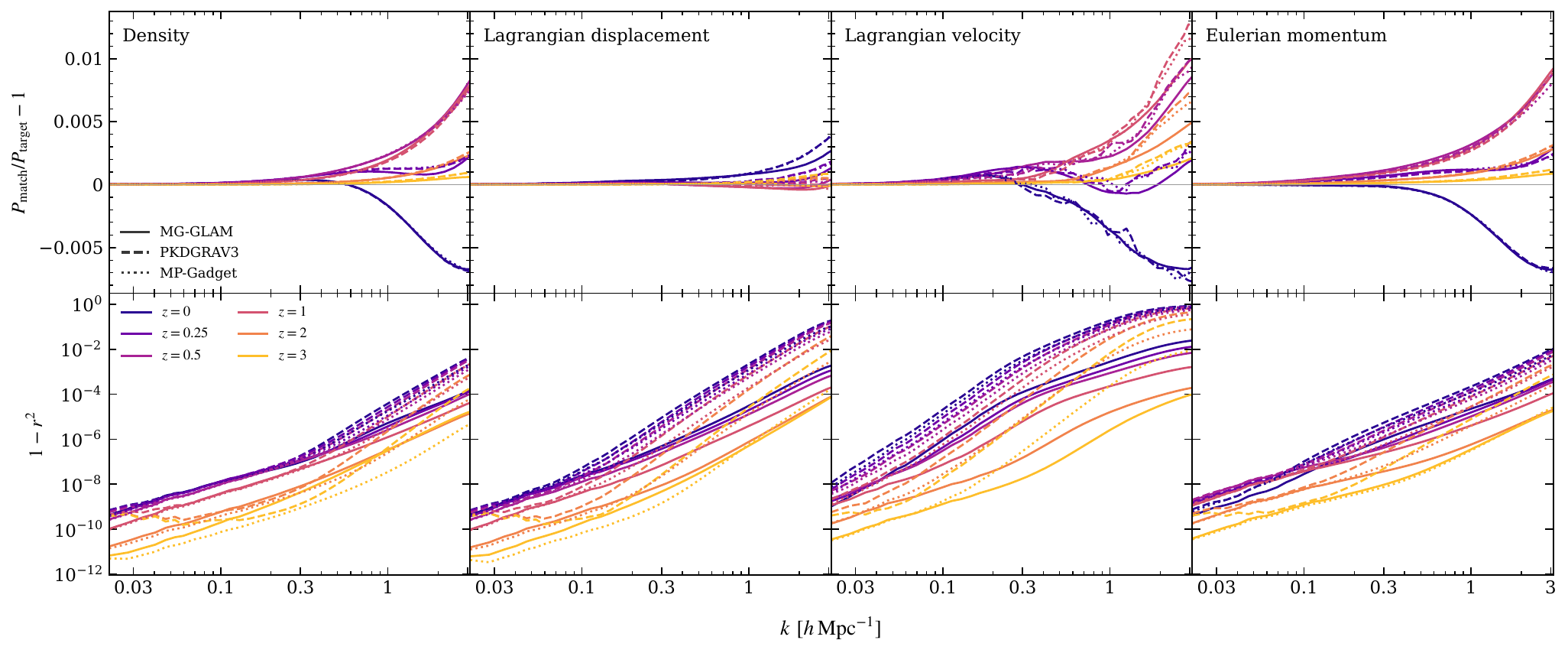}
    \caption{Power spectrum error $P_{\rm match}/P_{\rm target}-1$ (top) and stochasticity $1-r^2$ (bottom) of the density, Lagrangian displacement, Lagrangian velocity and Eulerian momentum for the DESI-like target, simulated with {\textsc{mg-glam}} (solid), {\textsc{pkdgrav3}} (dashed) and {\textsc{mp-gadget}} (dotted) from the same initial conditions with $1024^3$ particles. Each curve compares the target and match run with the same code, and the velocities are rescaled with Eq.~\eqref{eq:velcorr}. Eulerian fields are reproduced considerably better than individual particle positions and velocities, because particle dynamics inside haloes is chaotic. This is not specific to our method: at these scales, different $N$-body codes disagree with one another on the trajectories of individual particles inside haloes, even when started from the same initial conditions.}
    \label{fig:codes}
\end{figure*}

To test the extent to which our method is independent on the specific choice of $N$-body solver employed to produce the relevant $N$-body simulations, we simulate the DESI-like target and its match using the tree code \textsc{pkdgrav3}\footnote{\url{https://pkdgrav3.readthedocs.io}} \citep{Potter2017} and the TreePM code \textsc{mp-gadget}\footnote{\url{https://github.com/MP-Gadget/MP-Gadget}} \citep{Feng2018}. 

All three codes start from the same 2LPT initial conditions generated with \textsc{fml} at $z_{\rm init}=100$, with $1024^3$ particles in a box of side length $1\,h^{-1}{\rm Gpc}$, and use the same \ac{CPL} background without radiation. \textsc{pkdgrav3} computes the forces with the fast multipole method; following \citet{Potter2017}, we choose an opening angle of 0.4--0.7 and a gravitational softening of 1/50 of the mean interparticle separation ($19.5\,h^{-1}{\rm kpc}$), and integrate the particles with individual time steps over 600 base steps. For \textsc{mp-gadget}, following \citet{Bird2022}, we compute the long-range forces on a $2048^3$ mesh, and the short-range forces with a softening of $1/30$ of the mean interparticle separation, $32.6\,h^{-1}{\rm kpc}$, and integrate the particles with individual time steps over about 570 base steps. The force resolution of \textsc{mg-glam} is instead set by its mesh cell size of $244\,h^{-1}{\rm kpc}$, 7.5 and 12.5 times larger than the softening of \textsc{mp-gadget} and \textsc{pkdgrav3}. All statistics are measured as in Sect.~\ref{sec:results}, and the velocity rescaling of Eq.~\eqref{eq:velcorr} is applied in all three codes. 

For each code, we compare its own target and matched simulations in Fig.~\ref{fig:codes} (top). In the bins of Fig.~\ref{fig:codes}, the power spectrum errors of the three codes agree to within 0.1\% for $k\lesssim1\,h\,{\rm Mpc}^{-1}$ and 0.3\% up to near the Nyquist frequency, for all four fields and at all six redshifts. For the density and the momentum, they agree to within 0.04\% and 0.08\%. These differences are small compared with the matching errors themselves, which reach 0.9\% for the density near the Nyquist frequency. Differences in the absolute power spectra of the codes, for example from the force resolution, largely cancel in the target/match ratios, so the power spectrum error of the matching does not depend on the $N$-body solver. 

The stochasticity, however, differs substantially between the codes (Fig.~\ref{fig:codes}, bottom). On large scales, $k\lesssim0.1\,h\,{\rm Mpc}^{-1}$, the three codes agree, with a stochasticity of the density below $2\times10^{-8}$ in all of them. Towards higher $k$ and lower redshift, it grows faster in \textsc{mp-gadget} and \textsc{pkdgrav3} compared to \textsc{mg-glam}. At $z=0$ and $k\lesssim1\,h\,{\rm Mpc}^{-1}$, the stochasticity reaches $3.8\times10^{-5}$ for the density and $2.2\times10^{-4}$ for the momentum in \textsc{pkdgrav3}, and $1.8\times10^{-5}$ and $1.2\times10^{-4}$ in \textsc{mp-gadget}, compared to $5.2\times10^{-6}$ and $2.2\times10^{-5}$ in \textsc{mg-glam}. For the Lagrangian displacement and velocity, it reaches $2.2\times10^{-3}$ and 0.19 in \textsc{pkdgrav3} and $1.2\times10^{-3}$ and 0.14 in \textsc{mp-gadget}, compared with $3.1\times10^{-5}$ and $2.8\times10^{-3}$ in \textsc{mg-glam}. The Lagrangian fields are thus more affected than the Eulerian fields. At $z=0$ and $k\lesssim1\,h\,{\rm Mpc}^{-1}$, their stochasticity in the two tree codes is 1.5--2 orders of magnitude higher than in \textsc{mg-glam}, compared with 0.5--1 orders of magnitude for the density and the momentum. The  Lagrangian velocity is almost uncorrelated on small scales, with a stochasticity of 0.84 and 0.72 near the Nyquist frequency in \textsc{pkdgrav3} and \textsc{mp-gadget}.

We attribute this difference to the $N$-body codes rather than to the matching method which, as we mentioned above, is exact for identical $a(\tau)$ (Sect.~\ref{sec:matching}). $N$-body dynamics is chaotic on cluster scales: small differences between two runs, here caused by the slightly different expansion histories and initial conditions, grow quickly in dense regions, more so as the softening decreases \citep{Suto1991,Hayes2003,Genel2019,Asano2026}. Inside haloes, the trajectories of individual particles in the target and match therefore decorrelate, while the distribution of mass, and hence the Eulerian fields, remain much more similar \citep{Thiebaut2008,ElZant2019}. Individual particle displacement and velocities are most affected, in particular the Lagrangian velocity, which on small scales is dominated by the orbital motion of particles in haloes. The PM force of \textsc{mg-glam}, with a resolution of one mesh cell size, smooths the potential inside haloes and suppresses this chaotic growth. Its low stochasticity, especially for the Lagrangian fields, therefore partly reflects this limited force resolution. The Eulerian fields, in contrast, remain well correlated in all three codes: even at $z=0$, where the stochasticity is largest, it stays below $4\times10^{-5}$ for the density and $3\times10^{-4}$ for the momentum for $k\lesssim1\,h\,{\rm Mpc}^{-1}$. 

Since observables such as the galaxy clustering and redshift-space distortions are measured in Eulerian space, the Eulerian fields are the most relevant for cosmological applications. For them, the power spectrum error does not depend on the code, and the stochasticity remains small in all three codes.

\section{Velocity rescaling}
\label{app:velcor}

\begin{figure*}
    \centering
    \includegraphics[width=1\linewidth]{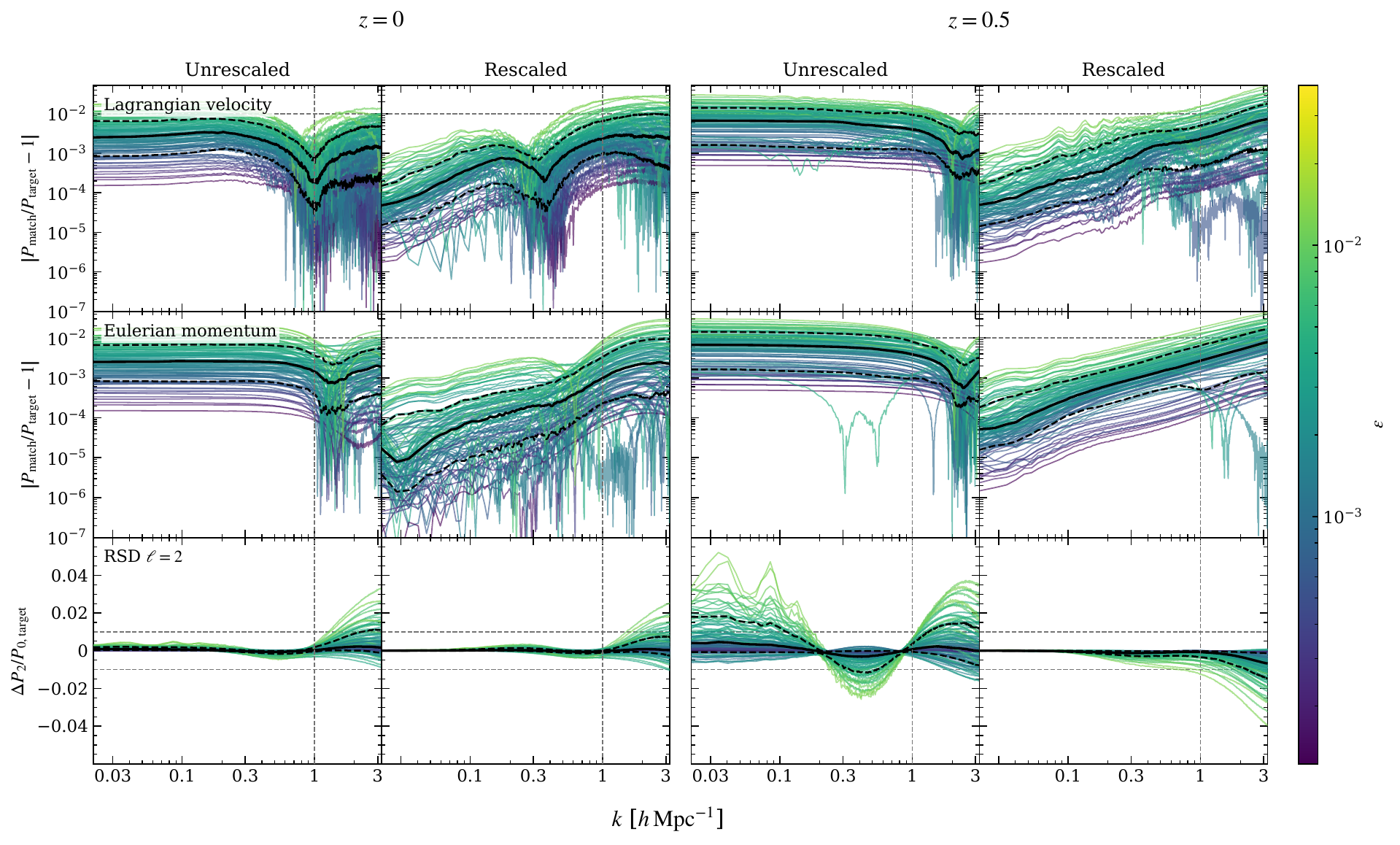}
    \caption{Effect of the velocity rescaling at $z=0$ (left two columns) and $z=0.5$ (right two columns), before (first column of each pair) and after (second column)rescaling. Top and middle: power spectrum error $|\Delta P/P|$ of the Lagrangian velocity and Eulerian momentum. Bottom: residual of the redshift-space quadrupole, $\Delta P_2/P_{0,{\rm target}}$. Colours and lines as in Fig.~\ref{fig:ensemble}. Grey dashed lines in the bottom row mark $\pm1\%$ error.}
    \label{fig:velcor}
\end{figure*}

Figure~\ref{fig:velcor} shows the effect of the velocity rescaling Eq.~\eqref{eq:velcorr} at $z=0$ and $0.5$. 

Without the rescaling, the power spectrum error of the Lagrangian velocity stays roughly a constant at $k\lesssim0.1\,h\,{\rm Mpc}^{-1}$, with a median of 0.28\% at $z=0$ and $-0.67\%$ at $z=0.5$, following the linear prediction of $2\Delta\ln(aHf)$. The rescaling removes this constant offset and reduces the median error on large scales by one to two orders of magnitude. The momentum behaves in the same way. 

At $k\gtrsim1\,h\,{\rm Mpc}^{-1}$, the remaining error, with a median of a few tenths of a percent, arises from nonlinear evolution and so cannot be removed by a constant rescaling. In fact, the error on these scales has the opposite sign (hence the dip at $k\simeq1\,h\,{\rm Mpc}^{-1}$ in the log-scale plot of Fig.~\ref{fig:velcor}), so the rescaling shifts the zero-crossing of the error to larger scales. This is also why, on nonlinear scales, the rescaled case has a slightly larger error than the unrescaled one. Because the rescaling is by a constant factor, the stochasticity of the velocity and momentum is unchanged.

The rescaling also affects the redshift-space distortions, because it enters the redshift-space positions through the velocities (Sect.~\ref{sec:matching}). The bottom row of Fig.~\ref{fig:velcor} shows the residual of the quadrupole, normalised by the target monopole as in Sect.~\ref{sec:fields}. At $z=0.5$ without the rescaling, the 16th--84th percentile range of the residuals is $-0.1\%$ to 1.8\% on large scales ($k=0.03\,h\,{\rm Mpc}^{-1}$), and the residuals of the models with the largest $\epsilon$ reach $\sim5\%$. After the rescaling, the residuals of all models stay below 0.5\% for $k\le0.2\,h\,{\rm Mpc}^{-1}$ and below 1.3\% for $k\lesssim1\,h\,{\rm Mpc}^{-1}$, as in Fig.~\ref{fig:rsd}. At $z=0$, where the velocity offset is smaller, the residual is below 0.5\% for $k\lesssim1\,h\,{\rm Mpc}^{-1}$ both before and after the rescaling. The rescaling is therefore needed for accurate redshift-space statistics at intermediate redshift, where the quantity $aHf$, as computed on the target and match models, differs the most.


\bsp	
\label{lastpage}
\end{document}